\documentclass[pdflatex,sn-nature,iicol]{sn-jnl}

\usepackage{graphicx}%
\usepackage{multirow}%
\usepackage{amsmath,amssymb,amsfonts}%
\usepackage{amsthm}%
\usepackage{mathrsfs}%
\usepackage[title]{appendix}%
\usepackage{xcolor}%
\usepackage{textcomp}%
\usepackage{manyfoot}%
\usepackage{booktabs}%
\usepackage{algorithm}%
\usepackage{algorithmicx}%
\usepackage{algpseudocode}%
\usepackage{listings}%
\usepackage{comment}
\usepackage{braket}
\makeatletter 
\renewcommand\@biblabel[1]{#1.} 
\makeatother

\begin{document}

\title[Article Title]{Decay-protected superconducting qubit with fast control enabled by integrated on-chip filters}

\author*[1]{\fnm{Aashish} \sur{Sah}}\email{aashish.sah@aalto.fi}
\author[1]{\fnm{Suman} \sur{Kundu}}
\author[1]{\fnm{Heikki} \sur{Suominen}}
\author[1]{\fnm{Qiming} \sur{Chen}}
\author[1,2]{\fnm{Mikko} \sur{M\"ott\"onen}}

\affil[1]{\orgdiv{QCD Labs, QTF Centre of Excellence, Department of Applied Physics}, \orgname{Aalto University}, \orgaddress{\street{Tietotie 3}, \city{Espoo}, \postcode{02150}, \country{Finland}}}

\affil[2]{\orgname{VTT Technical Research Centre of Finland Ltd}, \orgaddress{\street{Tietotie 3}, \city{Espoo}, \postcode{02150}, \country{Finland}}}

\abstract{ 
Achieving fast gates and long coherence times for superconducting qubits presents challenges, typically requiring either a stronger coupling of the drive line or an excessively strong microwave signal to the qubit. To address this, we introduce on-chip filters of the qubit drive exhibiting a stopband at the qubit frequency, thus enabling long coherence times and strong coupling at the subharmonic frequency, facilitating fast single-qubit gates, and reduced thermal load. The filters exhibit an extrinsic relaxation time of a few seconds while enabling sub-10-ns gates with subharmonic control. Here we show up to 200-fold improvement in the measured relaxation time at the stopband. Furthermore, we implement subharmonic driving of Rabi oscillations with a $\pi$~pulse duration of 12~ns. Our demonstration of on-chip filters and efficient subharmonic driving in a two-dimensional quantum processor paves the way for a scalable qubit architecture with reduced thermal load and noise from the control line.
}

\maketitle

\section*{Introduction}\label{Introduction}
The physical realization of a universal useful quantum computer comes with very stringent technical requirements, including those summarized in the DiVincenzo criteria~\cite{DiVincenzo2000}. One such requirement is a high coherence-time-to-gate-time (CT2GT) ratio, ensuring that the physical system representing a quantum bit, qubit, stores the encoded information with high fidelity until the execution of the assigned task. This property is reflected in the measured energy relaxation time of the qubit, $T_1$, which is the time scale for the qubit to exponentially lose an excitation and return to its ground state. Various qubit modalities such as trapped ions~\cite{Cirac1995, Bernien2017, Bluvstein2022, Bluvstein2023}, spin qubits~\cite{Loss1998,Xue2022, Weinstein2023, Burkard2023}, photonic qubits~\cite{Knill2001,Browne2005,OBrien2007,Shi2022}, and superconducting qubits~\cite{Nakamura1999, Koch2007, Devoret2005, Kockum2019} are being explored for longer relaxation times to enhance the CT2GT ratio.

Superconducting qubits, in particular, have exhibited significant progress, with relaxation times extending from a few nanoseconds~\cite{Nakamura1999, Mooij1999, Friedman2000, Vion2002, Han2001, Simmonds2004} to nearly a few milliseconds~\cite{Wang2022, Verjauw2022, Place2021, Somoroff2023} over the past two decades. The quest for further improvement of the coherence time is still ongoing. Alternatively, gates can be sped up to further improve the CT2GT ratio~\cite{Werninghaus2021, Ding2023}. However, this requires strong external coupling of the qubit to the drive line, which inadvertently decreases the coherence of the qubit, thereby posing a significant trade-off.


The coherence times of a qubit are influenced not only by the external coupling but also by intrinsic losses which in properly designed qubits, represent the primary loss mechanism, thus limiting the qubit performance. These losses arise from various sources, such as the presence of two-level-system defects near the qubit~\cite{Martinis2005, Gao2008, Lindstrom2009, Macha2010, Pappas2011, Neill2013, Muller2015, Wang2015} associated with dielectric losses and quasiparticle tunneling~\cite{Lutchyn2006, Paik2011, Catelani2011, Catelani2012}. Significant progress has been made in understanding~\cite{Paladino2014, Burnett2014, Pop2014, Lisenfeld2019, Yan2016, Vepsalainen2020, Murray2021, Siddiqi2021, Carroll2022} and mitigating these internal losses through the implementation of high-coherence materials~\cite{Wang2009, Megrant2012, Chang2013, Stern2014, Place2021, Somoroff2023}, fabrication recipes~\cite{Bruno2015, Wang2022, Verjauw2022, Deng2023}, and improved qubit geometry~\cite{He2022, Martinis2022}, leading to a situation where the external coupling should be made very weak not to limit qubit coherence, thus calling for high drive power at the qubit. 

While engineering very weak coupling to the qubit can increase the CT2GT ratio, it will impose a significant heat load on the dilution refrigerators that house the qubit, and especially on the attenuators that provide an electromagnetic environment for the qubit. To minimize the corresponding thermal noise reaching the qubit, attenuators are placed at multiple stages of the dilution refrigerators~\cite{Krinner}. Additionally, this will also limit the number of gate operations that can be carried out before the cooling capacity of the dilution refrigerator is exceeded.

Recent progress~\cite{Kono2020,Xia2023} has shown promise in addressing the above-discussed challenge of simultaneously achieving fast control and long coherence time. However, the approach of Ref.~\cite{Kono2020} raises concerns of quasiparticle generation due to the saturation of the Josephson junction employed in the qubit drive line at high power levels. Another important work~\cite{Xia2023} utilizes a novel way to drive a three-dimensional (3D) transmon qubit by subharmonics of the qubit resonance. Namely, the non-linearity of the qubit up-converts the drive at one-third of the qubit frequency to the qubit frequency in the spirit of three-wave mixing~\cite{Xia2023}. Inconveniently for scaling, the previous work employs bulky low-pass filters in the control line to suppress the spontaneous qubit decay to the strongly coupled drive line. Furthermore, to enable subharmonic control, only 43~dB of attenuation is used in the control line, which leads to a significant thermal noise photon of around 0.145 at qubit frequency of 5~GHz. 

In this work, we introduce designs and provide implementations of on-chip filters for subharmonic driving in two-dimensional qubit schemes. The on-chip filters are devised to fully isolate the qubit from the drive line at its resonance frequency while establishing two orders of magnitude stronger coupling at the subharmonic frequency compared to standard drive lines. This result enables fast single-qubit gates via subharmonic drive while demonstrating the \ensuremath{T_1} limit owing to the external coupling of seconds, thereby substantially improving the corresponding CT2GT ratio. Encouragingly, the introduced on-chip filters facilitate the use of over 60~dB of attenuation in the control lines to ensure a thermal noise photon below $2.5 \times 10^{-3}$. Furthermore, we show that the use of a bulky low-pass filter creates unwanted resonances in the qubit drive line, modulating the measured \ensuremath{T_1} in extreme ranges. Thus our demonstration of subharmonic driving of a qubit in two dimensions, integrated with an on-chip filter, seems a promising pathway for scalable superconducting quantum processing units.

\section*{Results}\label{Results}

\begin{figure*}[ht]
\includegraphics[width=\linewidth]{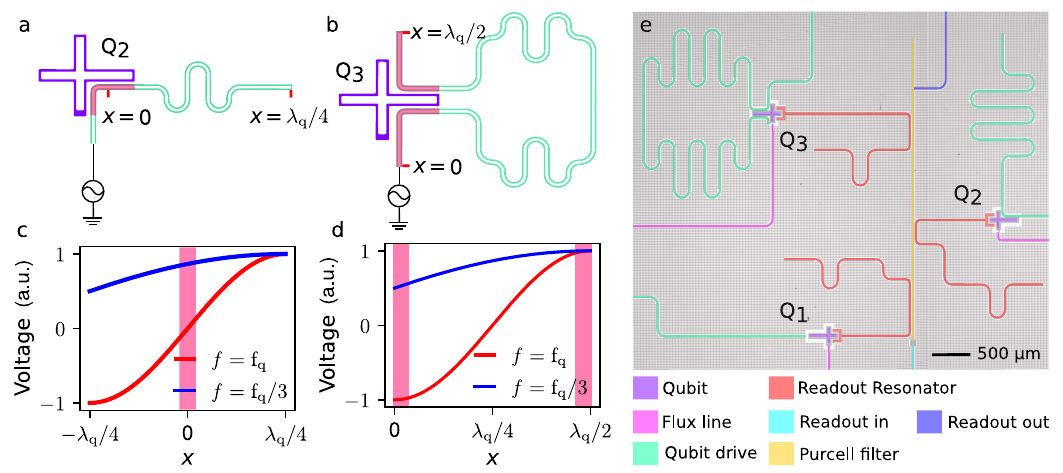}
\caption{\textbf{Device layout with integrated on-chip filters.} \textbf{a,b} Schematic representation of qubits (white crosses) strongly coupled to a transmission line (turquoise) implemented as (\textbf{a}) \ensuremath{\mathrm{\lambda/4}}  and (\textbf{b}) \ensuremath{\mathrm{\lambda/2}} filters at the qubit frequency \ensuremath{\mathrm{f_q}} which corresponds to the full wavelength \ensuremath{\lambda_\mathrm{q}}. \textbf{c,d} Voltage amplitude of the transmission line mode as a function of position \ensuremath{x} along the transmission line for the (\textbf{c}) \ensuremath{\mathrm{\lambda/4}} and (\textbf{d}) \ensuremath{\mathrm{\lambda/2}} filters. At the resonance frequency \ensuremath{f = \mathrm{f_q}} (red line), both on-chip filters effectively decouple from the qubit due to net zero voltage at the qubit, whereas at the subharmonic frequency \ensuremath{f = \mathrm{f_q/3}} (blue line), the on-chip filters couple strongly to the qubit. The pink shading highlights the region where the modes couple to the qubit. \textbf{e} False-colour optical-microscope image of the device. The device consists of three flux-tunable transmon qubits (violet) labeled Q1, Q2, and Q3. Each qubit is individually coupled to its own readout resonator (red) and flux lines (magenta). The readout resonators are further coupled to a common Purcell filter (yellow) for qubit readout. Additionally, qubits Q2 and Q3 are coupled to \ensuremath{\mathrm{\lambda/4}}  and \ensuremath{\mathrm{\lambda/2}} filters respectively, whereas qubit Q1 is weakly coupled to a standard transmission line for control (turquoise).
}
\label{fig:sample_layout}
\end{figure*}

\subsection*{Design and implementation of on-chip filters}

We have developed two distinct versions of on-chip filters: \ensuremath{\mathrm{\lambda/4}} and \ensuremath{\mathrm{\lambda/2}} filters as shown in Fig.~\ref{fig:sample_layout}a,~b. Essentially, these filters are coplanar-waveguide (CPW) transmission lines that connect capacitively to the qubit at the position \ensuremath{x=0} and end in an open circuit at \ensuremath{x=L_\mathrm{f}}, where \ensuremath{L_\mathrm{f}} is the length of the filter. In the case of the \ensuremath{\mathrm{\lambda/4}} filter, \ensuremath{L_\mathrm{f} = \lambda_\mathrm{q}/4}, where \ensuremath{\lambda_\mathrm{q}} is the wavelength of the mode in the drive line at the qubit frequency. Similarly, for the \ensuremath{\mathrm{\lambda/2}} filter, \ensuremath{L_\mathrm{f} = \lambda_\mathrm{q}/2}. We also note that \ensuremath{\mathrm{\lambda/2}} filter couples to the qubit at two separate positions as opposed to the single-point coupling in the case of the \ensuremath{\mathrm{\lambda/4}} filter. This two-point coupling design is inspired from multi-point coupling investigated in waveguide quantum electrodynamics \cite{FriskKockum2014, Kannan2020}.

\begin{figure*}[ht]
\includegraphics[width=\linewidth]{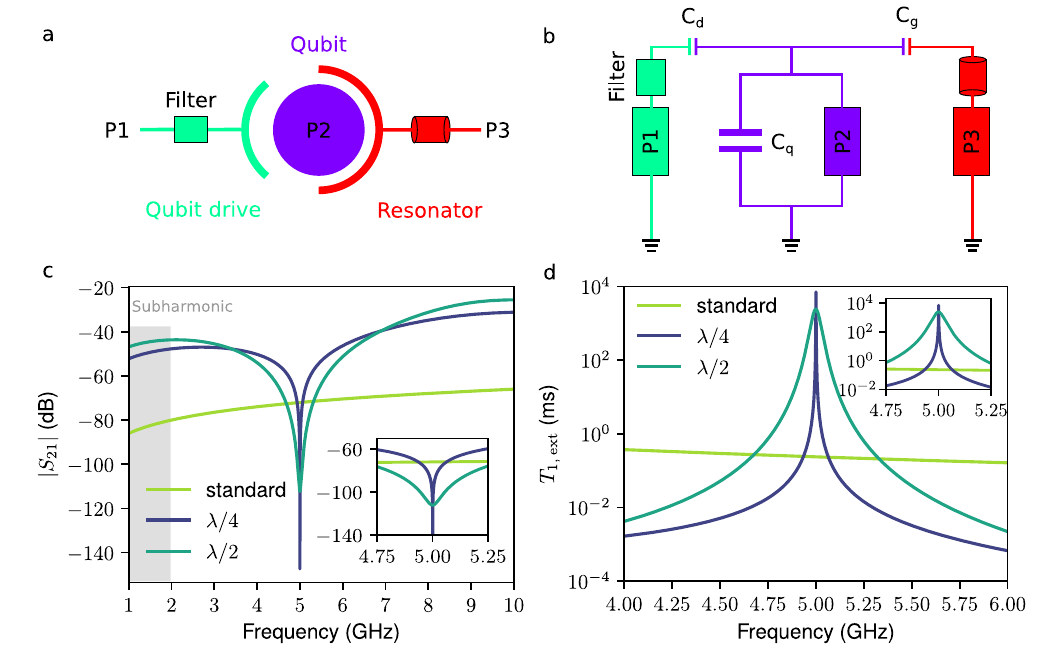}
\caption{\textbf{Simulated transmission and relaxation time of the qubit from drive lines.} \textbf{a} Schematic of a 3-port network labeled P1, P2, and P3. Port P1 is connected to a qubit drive line through a filter which is a standard transmission line for Q1,  \ensuremath{\lambda/4} filter for Q2, and \ensuremath{\lambda/2} filter for Q3. \textbf{b} Equivalent lumped-element model, where \ensuremath{\mathrm{C_d}} is the capacitance between the qubit and the drive line, \ensuremath{\mathrm{C_g}} is the capacitance between the qubit and the readout resonator, and \ensuremath{\mathrm{C_q}} is the shunt capacitance of the qubit to the ground. The coupling capacitances \ensuremath{\mathrm{C_d}} are 80~aF, 4.5~fF, and 9.2~fF for qubits Q1, Q2, and Q3, respectively. \textbf{c} Transmission amplitude from the qubit port to the drive line, \ensuremath{S_\mathrm{21}}, as a function of frequency for the different filter configurations as indicated. The \ensuremath{\lambda/4} and \ensuremath{\lambda/2} filters are designed to have a stopband at 5~GHz. The inset highlights the region in the vicinity of the filter stopband for convenient comparison. The shaded gray region corresponds to transmission at subharmonic frequency. \textbf{d} Estimated relaxation time, \ensuremath{T_\mathrm{1, ext}}, of the qubit owing to its coupling just to the drive line as a function of qubit frequency. This estimation is based on the total capacitance of the qubit and the input admittance \ensuremath{Y_\mathrm{in}} parallel to the qubit port P2. At the stopband, the estimated relaxation times of the qubit for the standard drive, the \ensuremath{\lambda/4} filter, and for the \ensuremath{\lambda/2} filter are approximately 0.3~ms, 7161~ms, and 2421~ms, respectively.} 
\label{fig:transmission_relaxation_numerical}
\end{figure*}
The working principle of the on-chip filters is illustrated in Fig.~\ref{fig:sample_layout}c,~d. In the \ensuremath{\mathrm{\lambda/4}} filter, the open-end creates a boundary condition for the voltage along the transmission line, resulting in a voltage anti-node at the open-end and a node at the qubit location when operating at the resonance frequency of the qubit as depicted in Fig.~\ref{fig:sample_layout}c. In contrast, the \ensuremath{\mathrm{\lambda/2}} filter exhibits a voltage anti-node at both the \ensuremath{x=0} and \ensuremath{x=\lambda/2} locations, but with opposite polarities as shown in Fig.~\ref{fig:sample_layout}d. Both configurations enforce net-zero voltage amplitude for the qubit, effectively decoupling the qubit from the drive line at the qubit resonance frequency. However, at the subharmonic frequency, the voltage profile at the qubit is close to its maximum, thereby establishing a strong coupling between the qubit and the drive line. The length, over which the mode voltage is integrated to obtain the effective coupling strength between the mode and the qubit, is highlighted in Fig.~\ref{fig:sample_layout}a-d. 

Our realized device is shown in Fig.~\ref{fig:sample_layout}e. The on-chip filters are integrated as a part of the qubit drive line. The device features three flux-tunable Xmon-style transmon qubits, labeled Q1, Q2, and Q3~\cite{Barends2013}. Qubit Q1 is connected to a standard, weakly coupled CPW transmission line. In contrast, qubits Q2 and Q3 are connected to \ensuremath{\mathrm{\lambda/4}}  and \ensuremath{\mathrm{\lambda/2}} filters, respectively. Additionally, the device includes flux lines to tune the frequencies of the qubits. Each qubit is coupled to its respective quarter-wavelength resonator for readout. These readout resonators are inductively and capacitively connected to a common quarter-wavelength Purcell filter~\cite{Jeffrey2014} to suppress qubit decay through the resonator.

We target the length of the on-chip filters such that their stopband is centered around 5~GHz. The results of our numerical simulations of these filters are shown in Fig.~\ref{fig:transmission_relaxation_numerical}. For a detailed description of the electromagnetic (EM) simulations, we refer to the Supplementary Note~S1 in Supplementary Information.

Figures~\ref{fig:transmission_relaxation_numerical}a,~b depict a schematic and its equivalent lumped-element model of a qubit coupled to an on-chip filter and to a readout resonator. In the model, 50-\ensuremath{\Omega} lumped ports, labeled P1 for the drive line, P2 for the qubit, and P3 for the readout resonator, are introduced for EM simulations. We focus on the scattering parameter \ensuremath{S_\mathrm{21}}, which relates to the power transmission from the drive line to the qubit port, and on the input admittance \ensuremath{Y_\mathrm{in}} parallel to the qubit. Here, we are only interested in the coupling between the drive line and the qubit. Therefore, both $S_{21}$ and \ensuremath{Y_\mathrm{in}} are calculated for the drive line port P1 matched to an impedance of 50~\ensuremath{\Omega} and the resonator port P3 grounded. With this procedure, one can estimate the external coupling rate just from the drive line to be \ensuremath{\gamma_\mathrm{q}(\omega) = \mathrm{Re}[Y_\mathrm{in}(\omega)]/C_\mathrm{q}}~\cite{Esteve1986, Houck2008a}. In the low-temperature limit, the relaxation time is given by \ensuremath{T_\mathrm{1}(\omega) = [1/T_\mathrm{1, ext}(\omega) + 1/T_\mathrm{1, diel}(\omega) + 1/T_\mathrm{1, oth}(\omega)]^{-1}}, where \ensuremath{T_\mathrm{1, ext}(\omega) = 1/\gamma_\mathrm{q}(\omega)} is the contribution from just the external coupling to the drive line, \ensuremath{T_\mathrm{1, diel}(\omega)} corresponds to the dielectric losses of the qubit, and \ensuremath{T_\mathrm{1, oth}(\omega)} accounts for other sources of relaxation such as readout resonators, flux lines, etc.

Figure~\ref{fig:transmission_relaxation_numerical}c shows that within the filter stopband centered at 5~GHz, the transmission from the qubit port to the transmission line lies below $-110$~dB, greatly suppressed from $-72$~dB obtained for the standard drive line. Importantly, within the subharmonic frequency range of 1--2~GHz, we observe an increase of over 30~dB in the transmission of the on-chip filters over the standard case.

In Fig. \ref{fig:transmission_relaxation_numerical}d, we examine the relaxation time \ensuremath{T_\mathrm{1, ext}} of the qubit resulting from the external coupling to the drive line. Within the filter stopband, we observe an enhancement in \ensuremath{T_\mathrm{1, ext}} by a factor of over 1000 with the \ensuremath{\lambda/4}  and \ensuremath{\lambda/2} filters compared with the standard drive line. The estimated \ensuremath{T_\mathrm{1, ext}} times at 5~GHz are as follows: 0.3~ms for the standard drive, 7161~ms for \ensuremath{\lambda/4} filter, and ~2421~ms for \ensuremath{\lambda/2} filter. The widths of the frequency ranges where qubit \ensuremath{T_\mathrm{1, ext}} exceeds 1~ms are 70~MHz and 450~MHz for \ensuremath{\lambda/4}  and \ensuremath{\lambda/2} filter, respectively. We attribute this difference of feasible frequency ranges to the fact that the \ensuremath{\lambda/2} filter is of higher order than the \ensuremath{\lambda/4} filter since the coupling points of the qubit are located at the voltage anti-nodes where the gradient of the voltage with respect to position vanishes.

Assuming a coupling that results in \ensuremath{T_\mathrm{1, ext}=1}~ms for a qubit driven resonantly through a standard transmission line, we estimate that a microwave pulse with a peak power of $-11$~dBm at room temperature is necessary to implement 10-ns single-qubit gates. With the above simulated parameters of the on-chip filters, corresponding peak powers of $-16$~dBm and $-22$~dBm will be required for \ensuremath{\lambda/4} filter and \ensuremath{\lambda/2} filter, respectively, using a subharmonic drive approach. Thus the filters combined with subharmonic driving may yield a significant reduction in power consumption for single-qubit gates. These values are calculated considering 60~dB of attenuation in the drive line and a simple rectangular pulse shape. Note that the peak powers may significantly increase for complex pulse shapes, which are currently being used to minimize the leakage out of the computational subspace for high-fidelity gates~\cite{Werninghaus2021}.

Furthermore, with the above estimates, we obtain a heat dissipation of $-53$~dBm at the base plate resulting from a resonant drive with the standard transmission line. In the case of a subharmonic drive with the on-chip filters, we estimate heat dissipations of $-58$~dBm with \ensuremath{\lambda/4} filter and $-64$~dBm with \ensuremath{\lambda/2} filters.

\subsection*{Characterization of on-chip filters}

\begin{figure*}[ht]
\includegraphics[width=\linewidth]{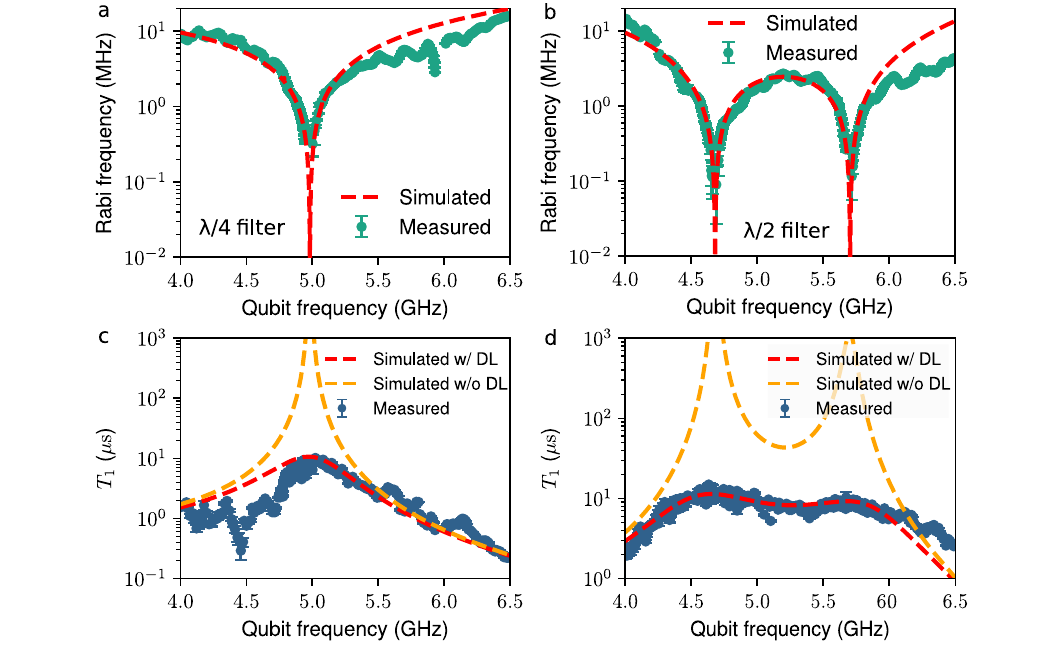}
\caption{\textbf{Characterization of the on-chip filters.} \textbf{a,b} Measured (dots) and simulated (dashed line) Rabi frequencies of the qubit for (\textbf{a}) \ensuremath{\lambda/4}  and (\textbf{b}) \ensuremath{\lambda/2} filters as functions of the qubit frequency \ensuremath{f_\mathrm{q}=\omega_\mathrm{q}/(2\pi)}. We swept qubit frequency in the range of 4 to 6.5~GHz and carried out the Rabi measurement to characterize the coupling strength of the drive lines to the qubit. The dashed line represents the Rabi frequency estimated using \ensuremath{f_\mathrm{R} = \Omega_\mathrm{R}/2\pi =  2\sqrt{\gamma_\mathrm{q}} \beta}, where \ensuremath{\gamma_\mathrm{q} = \mathrm{Re}[Y_\mathrm{in}]/C_\mathrm{q}} and \ensuremath{\beta\propto\omega_\mathrm{q}^{-1/2}V}, where V is the peak-to-peak voltage amplitude at the chip level. Note that the measured Rabi frequency for the \ensuremath{\lambda/2} filter shows two distinct stopbands centered around 4.7~GHz and 5.7~GHz, the presence of which is attributed to the unequal coupling at positions \ensuremath{x = 0} and \ensuremath{x = \mathrm{\lambda/2}}. \textbf{c,d} Measured (dots) and simulated (dashed line) relaxation time, \ensuremath{T_1}, of the qubit for (\textbf{c}) \ensuremath{\lambda/4}  and (\textbf{d}) \ensuremath{\lambda/2} filters as functions of the qubit frequency. The dielectric loss tangent of \ensuremath{3\times 10^{-6}} obtained from the fit is used to account for dielectric losses limiting the \ensuremath{T_1}, represented by the red dashed line. While orange dashed line corresponds to the case without the dielectric losses. The simulated $T_1$ also takes Purcell decay through the resonator into account. The error bars represent \ensuremath{1\sigma} fitting uncertainty.}
\label{fig:filter_characterization}
\end{figure*}

We follow the typical steps to fabricate the device on a silicon wafer, which includes a niobium metallization layer on top. The micron-sized features in the niobium are created using maskless lithography and etching, while the Manhattan-style Josephson junctions are patterned using electron-beam lithography and thermal evaporation of aluminium.  We cool down the device in a commercial Bluefors XLD500 dilution refrigerator for cryogenic measurements. For more detailed information on the fabrication process and the experimental setup, see the Supplementary Note~S2 \& S3 in Supplementary Information. 

We carry out typical time-domain measurements to characterize the device parameters summarized in Table S1 in Supplementary Information. To quantify the performance of the on-chip filters, we measured the Rabi frequency and the \ensuremath{T_1} of the qubit around the target filter frequency of 5~GHz. A comprehensive description of the measurement sequence can be found in greater detail in the Supplementary Note~S4 in Supplementary Information.

Figures~\ref{fig:filter_characterization}a, b show the measured Rabi frequency for qubits driven through the \ensuremath{\mathrm{\lambda/4}}  and \ensuremath{\mathrm{\lambda/2}} filters. We observe an impressive 50-fold and 220-fold suppression in measured Rabi frequency at the stopbands of \ensuremath{\lambda/4} and \ensuremath{\lambda/2} filters, respectively, corresponding to 34~dB and 47~dB changes in $S_{21}$. The suppression factor is defined with respect to the maximum Rabi frequency observed in the qubit frequency range of 4--6.5~GHz. The simulated data depicted in Fig. \ref{fig:filter_characterization}a, b correspond to the Rabi frequency derived from the input admittance \ensuremath{Y_\mathrm{in}} obtained through EM simulations, see Methods. Evidently, the measured and simulated Rabi frequencies exhibit a high level of agreement. Some minor deviations observed can be attributed to parasitic resonance and imperfections stemming from coupling with the control lines.

Note the significant discrepancy in the measured Rabi frequency for the \ensuremath{\lambda/2} filter shown in Fig.~\ref{fig:filter_characterization}b in comparison to the simulated transmission depicted in Fig.~\ref{fig:transmission_relaxation_numerical}c. In contrast to a single stopband at 5~GHz, we observe two stopbands centered around 4.7~GHz and 5.7~GHz. As detailed in Supplementary Note~S1 in Supplementary Information, this splitting of the stopband mainly arises from the asymmetry in the coupling capacitance between the drive line and the qubit at the positions \ensuremath{x=0} and \ensuremath{x=\lambda/2}, in addition to the fact that the coupling region at \ensuremath{x=\lambda/2} is of finite length. Consequently, we show a good agreement between the measured and simulated results in Fig.~\ref{fig:filter_characterization}b by introducing an asymmetric coupling at the open-end of the \ensuremath{\mathrm{\lambda/2}} filter in the EM simulation. 

Figures~\ref{fig:filter_characterization}c, d depict the measured \ensuremath{T_1} for the \ensuremath{\mathrm{\lambda/4}} and \ensuremath{\mathrm{\lambda/2}} filter. In our device, we consistently observe a \ensuremath{T_1} ceiling of approximately 10~\ensuremath{\mu\mathrm{s}} across all qubits. To investigate the reason for this limited \ensuremath{T_1}, we fit the Purcell decay and dielectric-loss models to the measured \ensuremath{T_1} of qubit Q1 near its readout resonator frequency, see Supplementary Note~S5 in Supplementary Information. From the fit, we obtain a dielectric loss tangent of \ensuremath{3\times 10^{-6}}, confirming dielectric loss as the primary loss mechanism in our device. Consequently, we used the dielectric loss tangent of \ensuremath{3\times 10^{-6}} in the EM simulations, yielding the simulation results for qubits Q2 and Q3 in good agreement with the experiments. Without considering dielectric losses in the simulation, the extrapolated $T_1$ at the filter frequency reaches hundreds of milliseconds range. Due to the Purcell decay through the resonator, the simulated $T_1$ at the second stopband centered around 5.7~GHz shows only a few milliseconds.

For the qubit \ensuremath{T_1} measured at the stopband of the filter, we observed a remarkable 20-fold improvement for the \ensuremath{\mathrm{\lambda/4}} filter and an even more substantial 200-fold improvement for the \ensuremath{\mathrm{\lambda/2}} filter in comparison to the values measured at the flux sweet-spot of the qubit, see Supplementary Table S1 in Supplementary Information. This enhancement has the potential for further improvements through the use of low-loss dielectric materials and implementation of state-of-the-art fabrication processes for high coherence.

\subsection*{Resonant and subharmonic drive}

\begin{figure*}[ht]
\includegraphics[width=\linewidth]{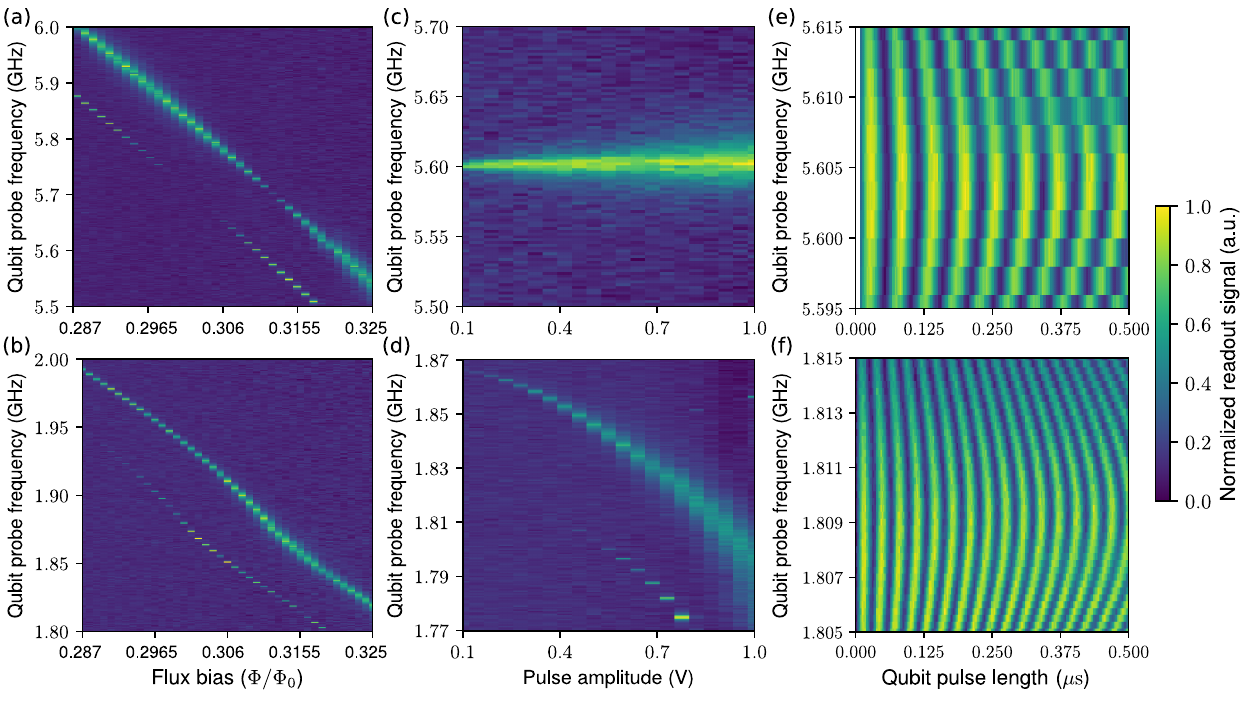}
\caption{\textbf{Spectroscopy and Rabi oscillations of qubit Q3 with resonant driving and subharmonic driving through the \ensuremath{\mathrm{\lambda/2}} filter}. \textbf{a,b} Normalized readout signal of the qubit as a function of the flux bias and the frequency of the (\textbf{a}) resonant drive and (\textbf{b}) subharmonic drive. The top diagonal feature indicates the lowest transition frequency of the qubit and the bottom feature reveals the two-photon transition to the second excited state of the qubit. \textbf{c,d} Normalized readout signal of the qubit as a function of the driving-voltage amplitude at room temperature and the frequency of the (\textbf{c}) resonant drive and (\textbf{d}) subharmonic drive for a flux bias of 0.32~$\Phi_\mathrm{0}$. Note the power broadening and up to $-70$~MHz alternating-current (AC) Stark shift in (\textbf{d}) with increasing pulse amplitude. \textbf{e,f} Normalized readout signal of the qubit as a function of the pulse length and the frequency of the (\textbf{e}) resonant drive and (\textbf{f}) subharmonic drive for a flux bias of 0.32~$\Phi_\mathrm{0}$. In (\textbf{e}), the pulse amplitude 1~V yields a Rabi frequency of 17.57~MHz and in (\textbf{f}), we have 0.9~V and 31.12~MHz, respectively, at the Stark-shifted frequency of 1.809~GHz. 
}
\label{fig:Resonant_subharmonic_driving}
\end{figure*}

Figures~\ref{fig:Resonant_subharmonic_driving}a,b present the flux spectroscopy results for qubit Q3 coupled to \ensuremath{\lambda/2} filter using both resonant and subharmonic driving techniques. By sweeping the flux bias \ensuremath{\Phi} threading through the SQUID in the units of flux quanta \ensuremath{\Phi_\mathrm{0}}, we tune the frequency of qubit Q3 in a range of 500~MHz near the second stopband of the \ensuremath{\lambda/2} filter centered around 5.7~GHz. With resonant driving, we observe a major reduction in the qubit linewidth as it approaches the stopband at 5.7~GHz. Note that also the two-photon process to excite the qubit from its ground state to the second excited state is visible at different flux bias but frequency matching to 5.7~GHz, consistent with the filter protecting the qubit at its stopband rather than the qubit itself being substantially different at a certain flux bias. However, with subharmonic driving, we observe a broadened qubit linewidth at around \ensuremath{f = 5.7/3 = 1.9}~GHz. This observation indicates that the \ensuremath{\lambda/2} filter exhibits strong coupling at the subharmonic frequency but very weak coupling at the resonance frequency, which agrees with our intended design and purpose.

Qubit power spectroscopy with resonant driving in Fig.~\ref{fig:Resonant_subharmonic_driving}c demonstrates a broadening of the qubit line  with increasing pulse amplitude. The spectroscopic line remains centered at the frequency of 5.6~GHz independent of the pulse amplitude. In contrast, Fig.~\ref{fig:Resonant_subharmonic_driving}d exhibits a pronounced AC Stark shift caused by the off-resonant nature of the subharmonic drive. The AC Stark shift of the qubit frequency is typical for an off-resonant driving of the qubit~\cite{Carroll2022}. This shift increases with increasing pulse amplitude, reaching around $-70$~MHz at the maximum amplitude used. The negative direction of the shift arises from the negative anharmonicity of the transmon qubit. 

Figures \ref{fig:Resonant_subharmonic_driving}e,f illustrate Rabi oscillations of the qubit under resonant and subharmonic driving. For room-temperature pulse amplitudes of 1~V and 0.9~V, the measured Rabi frequencies for resonant and subharmonic driving are 17.57~MHz and 31.12~MHz, respectively. In the case of subharmonic drive, the Rabi frequency corresponding to the slowest oscillation occurs with an AC Stark shift of $-57.67$~MHz.

\subsection*{Rabi frequency and AC Stark shift}

\begin{figure}[ht]
\includegraphics[width=\linewidth]{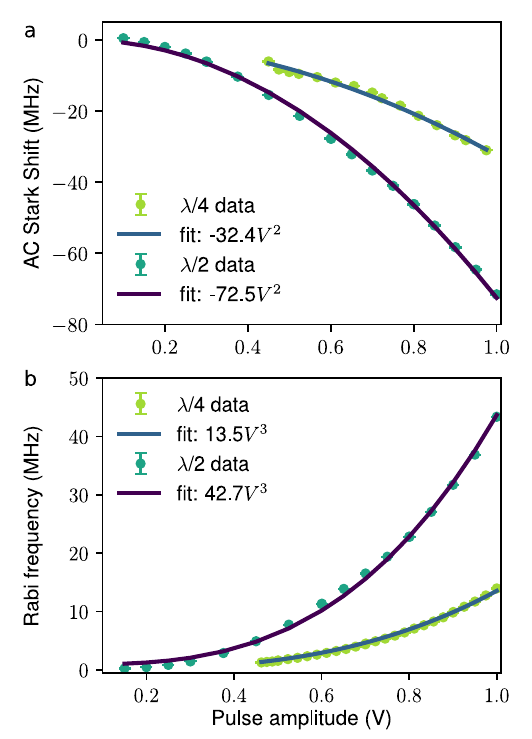}
\caption{\textbf{Stark shifts and Rabi frequencies in subharmonic driving of qubits Q2 and Q3.} \textbf{a,b} Measured (markers) and model-fitted (solid lines)  (\textbf{a}) AC Stark shift and (\textbf{b}) Rabi frequency for subharmonic driving through the \ensuremath{\lambda/4}  and \ensuremath{\lambda/2} filters as indicated as functions of the voltage amplitude of the drive pulse at room temperature. The error bars indicate \ensuremath{1\sigma} uncertainty.
}
\label{fig:Subharmonic_Stark_Rabi}
\end{figure}

Figure~\ref{fig:Subharmonic_Stark_Rabi}a shows the AC Stark shift arising from the off-resonant nature of the subharmonic drive. For a maximum pulse amplitude of 1~V, the measured shift is approximately $-33$~MHz for the \ensuremath{\lambda/4} filter and $-73$~MHz for the \ensuremath{\lambda/2} filter. We obtain a good agreement between the measured Stark shift and the fit using a quadratic function in the pulse amplitude as theoretically expected from equation \eqref{eq:subharmonic_H} in the Method section.

In Fig.~\ref{fig:Subharmonic_Stark_Rabi}b, we conduct a performance comparison between the \ensuremath{\lambda/4} and \ensuremath{\lambda/2} filters using the subharmonic drive. This comparison involves varying the pulse amplitude to measure the Rabi frequency at the AC-Stark-shifted frequency studied in Fig.~\ref{fig:Subharmonic_Stark_Rabi}a. We observe that the maximum measured Rabi frequency is approximately 13~MHz and 43~MHz, resulting in \ensuremath{\pi} pulses of lengths 37~ns and 12~ns with the \ensuremath{\lambda/4} and \ensuremath{\lambda/2} filters, respectively.  We employ a cubic function to fit the measured Rabi frequencies as a function of the pulse amplitude, showing an excellent agreement with the experimental results as expected from the three-wave-mixing nature of subharmonic driving, as can be seen from equation \eqref{eq:subharmonic_H} in the Method section. Importantly, we maintain over 70~dB of attenuation in the drive line to ensure that the thermal occupation of the qubit remains below \ensuremath{0.5\%}, and limit heat dissipation at the base plate to less than $-40$~dBm~$=100$~nW.

\section*{Discussion}\label{Discussion}

In this work, we addressed the seemingly competing challenges of implementing fast single-qubit gates and mitigating the decoherence of qubits due to the strong coupling needed for such fast gates. These challenges are intimately related to the power dissipation budget in the dilution refrigerator that houses the qubits and cross-talk mitigation between the qubits.

A transmon qubit with a standard capacitive coupling to its broadband drive line resulting in \ensuremath{T_\mathrm{1, ext}} of 1~ms requires a peak power of $-11$~dBm at room temperature to implement 10-ns-long single-qubit gates using a resonant drive. This power requirement is estimated considering 60~dB attenuation in the drive line to keep thermal qubit excitations low. In contrast, a subharmonic drive requires roughly $25$~dBm in this scenario with no drive line filter. Such drive power leads to heat dissipation of about $-17.5$~dBm at the base plate, unfortunately matching its cooling capacity of $-17$~dBm. 

Our proposed on-chip filters are designed to achieve a \ensuremath{T_\mathrm{1, ext}} bound of 7161~ms for the \ensuremath{\lambda/4} filter and 2421~ms for the \ensuremath{\lambda/2} filter at the stopband near the qubit transition frequency. We estimate peak powers of $-16$~dBm for the \ensuremath{\lambda/4} filter and $-22$~dBm for the \ensuremath{\lambda/2} filter at room temperature to implement 10-ns-long gates using subharmonic driving. With 60~dB of attenuation in the drive lines, the heat dissipation at the base plate is reduced to $-58$~dBm for the \ensuremath{\lambda/4} filter and to $-64$~dBm for the \ensuremath{\lambda/2} filter, enabling tens of thousands of simultaneous gate operations before reaching the cooling power of typical refrigerators.

Experimentally, we achieved an impressive 50-fold and 220-fold suppression in the measured Rabi frequency and a factor of 50 and 80 improvements in the measured \ensuremath{T_1} at the stopbands of the \ensuremath{\lambda/4} and \ensuremath{\lambda/2} filters, respectively. We measured a maximum \ensuremath{T_1} of  10~\ensuremath{\mu\mathrm{s}} across the device, mainly limited by dielectric losses. The simulated and measured data exhibited excellent agreement, from which we conclude that the on-chip filters function as desired. We note that using state-of-the-art fabrication processes for high coherence and low-loss dielectric materials may further enhance these results and allow for more thorough study of the effects of the filters on the qubit energy relaxation time.

In our subharmonic-drive experiments, we obtained qubit spectroscopy akin to those with resonant driving using standard two-tone measurements. We measured \ensuremath{\pi} pulse lengths of 37~ns and 12~ns for the \ensuremath{\lambda/4} and \ensuremath{\lambda/2} filters, respectively, using Rabi measurements. This result was obtained with 62~dB of attenuation distributed across various dilution stages, combined with an additional 12~dB of attenuation from filters and wires at room temperature and inside the refrigerator, yielding approximately 74~dB in total in the drive line. This configuration effectively reduced thermal noise photons to below $2.5 \times 10^{-3}$ at qubit frequency and around $0.03$ at subharmonic frequency. This is a prominent improvement compared to the work in Ref.~\cite{Xia2023}, where single-qubit gate of length 35~ns has been demonstrated with only 43~dB attenuation in the qubit control line and with the use of bulky off-chip low-pass filters.

The off-resonant nature of the subharmonic drive led to a substantial AC Stark shift of around $-73$~MHz at a Rabi frequency of about 43~MHz with the \ensuremath{\lambda/2} filter, significantly more pronounced than in resonant driving. As a result, the standard approach for implementing and characterizing single-qubit gates through Randomized Benchmarking may not be feasible as such, calling for special attention to phase correction between gates. In addition, minimization of leakage errors with such brief gates is also pivotal. Although the drag scheme effectively reduces leakage errors in resonant driving, a similar approach adapted for subharmonic driving needs to consider the large AC Stark shift. Addressing this issue stands as an interesting topic for future research.

\section*{Methods}\label{Methods}

\subsection*{Single-qubit control}\label{subsec:sing_qub_ctrl}
The Hamiltonian describing the control of the qubit has the following form
\begin{align}\label{eq:qubit_drive_H}
    \hat{H} = \hat{H}_\mathrm{qubit} + \hat{H}_\mathrm{drive}(t),
\end{align}
where the first term on the right is the Hamiltonian of the undriven qubits and the the second term is the drive Hamiltonian.
The Hamiltonian of the transmon qubit is given by 
\begin{align}
    \hat{H}_\mathrm{qubit} = \hbar\omega_\mathrm{q} \hat{b}^\dagger \hat{b} + \frac{\hbar\alpha}{2} \hat{b}^\dagger\hat{b}^\dagger\hat{b}\hat{b},
\end{align}
where \ensuremath{\omega_\mathrm{q}} is the angular frequency of the lowest qubit transition, \ensuremath{\alpha} is the anharmonicity, and \ensuremath{\hat{b}} and \ensuremath{\hat{b}^\dagger}  are the annihilation and creation operators acting on the qubit states, respectively.
The driving Hamiltonian assumes the form

\begin{equation}
    \hat{H}_\mathrm{drive}(t) = \hbar\Omega_\mathrm{R}e^{-i\omega_\mathrm{d}t - \varphi_\mathrm{d}} \hat{b}^\dagger + \mathrm{h.c.},
\end{equation}
where \ensuremath{\Omega_\mathrm{R}} is the Rabi angular frequency, and \ensuremath{\omega_\mathrm{d}} and \ensuremath{\varphi_\mathrm{d}} are the frequency and the phase of the microwave drive. To implement arbitrary single-qubit gates, the Rabi frequency and the phase of the drive are temporally controlled. 

In the case of resonant driving, we have \ensuremath{\omega_\mathrm{d} \approx \omega_\mathrm{q}}. By going into a frame rotating at \ensuremath{\omega_\mathrm{d}}, we obtain a simplified qubit-drive Hamiltonian with the detuning \ensuremath{\delta_\mathrm{res} = \omega_\mathrm{q} - \omega_\mathrm{d}}, 
\begin{align}
\begin{split}
    \hat{H}_\mathrm{res} &= \hbar\delta_\mathrm{res}\hat{b}^\dagger\hat{b} + \frac{\hbar\alpha}{2} \hat{b}^\dagger\hat{b}^\dagger\hat{b}\hat{b} \\
    &+ \hbar(\Omega_\mathrm{R}\hat{b}^\dagger +  \mathrm{h.c.}).
\end{split}
\end{align}
For the subharmonic driving, instead we have \ensuremath{\omega_\mathrm{d} = \omega_\mathrm{q}/3 + \delta_\mathrm{sub}} and following the derivation in~\cite{Xia2023}, we find 
\begin{equation}\label{eq:subharmonic_H}
\begin{split}
    \hat{H}_\mathrm{sub} &= \hbar(2\alpha|\eta|^2 - 3 \delta_\mathrm{sub})\hat{b}^\dagger\hat{b} + \frac{\hbar\alpha}{2} \hat{b}^\dagger\hat{b}^\dagger\hat{b}\hat{b} \\ 
    &+ \frac{\hbar\alpha}{3}(\eta^3\hat{b}^\dagger +  \mathrm{h.c.}),
\end{split}
\end{equation}
where \ensuremath{\eta=\frac{\Omega_\mathrm{R}(\omega_\mathrm{q} - \alpha)}{\omega^2_\mathrm{d} - (\omega_\mathrm{q} - \alpha)^2}} represents the strength of the subharmonic drive. From the Hamiltonian, we obtain the AC Stark shift \ensuremath{\delta_\mathrm{sub}=2\alpha|\eta|^2/3} resulting from the off-resonant drive and the subharmonic Rabi frequency \ensuremath{\Omega^\mathrm{sub}_\mathrm{R} = 2\alpha|\eta|^3/3}.

\subsection*{Relation of the external coupling strength and Rabi frequency with admittance}
In our experiments, we capacitively couple a CPW transmission line to the qubit for control. The interaction strength between the drive line and the qubit is given by~\cite{Blais2021a}

\begin{equation}\label{eq:external_coupling}
    \gamma_\mathrm{q} = 2e^2\bigg(\frac{C_\mathrm{d}}{C_\mathrm{q}}\bigg)^2\Bigg(\frac{E_\mathrm{J}}{2E_\mathrm{C}}\Bigg)^\frac{1}{2}\frac{Z_\mathrm{tml}\omega_\mathrm{q}}{\hbar},
\end{equation}
where $e$ is the elementary charge, $C_\mathrm{d}$ is the capacitance between the qubit and the drive line, $E_\mathrm{J}$ is the Josephson energy, $E_\mathrm{C}$ is the charging energy of the qubit, $Z_\mathrm{tml}$ is the characteristic impedance of the drive line, and $C_\mathrm{q}$ is the capacitance of the qubit island, respectively.

We express the Rabi frequency in terms of the interaction strength as 
\begin{equation}\label{eq:Rabi_freq}
    \Omega_\mathrm{R} = 2\sqrt{\gamma_\mathrm{q}} \beta,
\end{equation}
where \ensuremath{\beta = \big(\frac{1}{2\hbar\omega_\mathrm{q}Z_\mathrm{tml}}\big)^\frac{1}{2} V} and $V$ is the voltage amplitude of the drive~\cite{Blais2021a}. From equations \eqref{eq:external_coupling} and \eqref{eq:Rabi_freq}, we observe that for the resonant case, the Rabi frequency \ensuremath{\Omega_\mathrm{R} \propto V}, whereas for subharmonic drive, \ensuremath{\Omega^\mathrm{sub}_\mathrm{R} \propto V^3}. This cubic scaling with respect to the voltage amplitude renders the subharmonic drive favourable over resonant driving at high amplitudes or coupling strengths. 

The above simplified analytical equation may be used to obtain \ensuremath{\gamma_\mathrm{q}} of the qubit at weak coupling to the transmission line. In our case, we also consider the complete system of a qubit strongly coupled to an on-chip filter and to the readout resonator. Here, we resort to electromagnetic simulations to estimate the external coupling strength using \ensuremath{\gamma_\mathrm{q}(\omega) = \mathrm{Re}[Y_\mathrm{in}(\omega)]/C_\mathrm{q}}, where \ensuremath{Y_\mathrm{in}} is the admittance in parallel with the qubit obtained from the simulations~\cite{Esteve1986, Houck2008a}.

\subsection*{Electromagnetic simulations}
The device layout is created using KQCircuits, an open-source Python library, within the KLayout editor. Subsequently, this layout is exported to the electromagnetic-simulation environment of the Sonnet software. Owing to the relatively large size of the qubit-drive geometry and the associated complexity of the simulation, we simplify the model by replacing a part of the transmission line by a lumped port, which allows for building a simulation of the full device using a lumped-circuit simulator.

The scattering-parameter data is exported from Sonnet to Microwave Office AWR, where we construct and simulate a full-circuit model. We extract and export the input admittance \ensuremath{Y_\mathrm{in}} for further analysis in Python. A comprehensive description of the steps taken to achieve these simulation results is presented in the Supplementary Note~S1 in Supplementary Information.

\backmatter

\section*{Data Availability}

All relevant data and codes generating the figures in this article are available via Zenodo at \href{https://doi.org/10.5281/zenodo.11234842}{https://doi.org/doi/10.5281/zenodo.11234842} \cite{sah_2024_11234843}.

\bibliography{main}

\begin{thebibliography}{10}
\expandafter\ifx\csname url\endcsname\relax
  \def\url#1{\burl{#1}}\fi
\expandafter\ifx\csname urlprefix\endcsname\relax\def\urlprefix{URL }\fi
\providecommand{\bibinfo}[2]{#2}
\providecommand{\eprint}[2][]{\url{#2}}
\providecommand{\doi}[1]{\url{https://doi.org/#1}}
\bibcommenthead

\bibitem{DiVincenzo2000}
\bibinfo{author}{DiVincenzo, D.~P.}
\newblock \bibinfo{title}{{The physical implementation of quantum computation}}.
\newblock \emph{\bibinfo{journal}{Fortschritte der Physik}} \textbf{\bibinfo{volume}{48}}, \bibinfo{pages}{771--783} (\bibinfo{year}{2000}).
\newblock \urlprefix\url{https://onlinelibrary.wiley.com/doi/10.1002/1521-3978}.

\bibitem{Cirac1995}
\bibinfo{author}{Cirac, J.~I.} \& \bibinfo{author}{Zoller, P.}
\newblock \bibinfo{title}{{Quantum computations with cold trapped ions}}.
\newblock \emph{\bibinfo{journal}{Physical Review Letters}} \textbf{\bibinfo{volume}{74}}, \bibinfo{pages}{4091--4094} (\bibinfo{year}{1995}).
\newblock \urlprefix\url{https://journals.aps.org/prl/abstract/10.1103/PhysRevLett.74.4091}.

\bibitem{Bernien2017}
\bibinfo{author}{Bernien, H.} \emph{et~al.}
\newblock \bibinfo{title}{{Probing many-body dynamics on a 51-atom quantum simulator}}.
\newblock \emph{\bibinfo{journal}{Nature}} \textbf{\bibinfo{volume}{551}}, \bibinfo{pages}{579--584} (\bibinfo{year}{2017}).
\newblock \urlprefix\url{https://www.nature.com/articles/nature24622}.

\bibitem{Bluvstein2022}
\bibinfo{author}{Bluvstein, D.} \emph{et~al.}
\newblock \bibinfo{title}{{A quantum processor based on coherent transport of entangled atom arrays}}.
\newblock \emph{\bibinfo{journal}{Nature}} \textbf{\bibinfo{volume}{604}}, \bibinfo{pages}{451--456} (\bibinfo{year}{2022}).
\newblock \urlprefix\url{https://www.nature.com/articles/s41586-022-04592-6}.

\bibitem{Bluvstein2023}
\bibinfo{author}{Bluvstein, D.} \emph{et~al.}
\newblock \bibinfo{title}{{Logical quantum processor based on reconfigurable atom arrays}}.
\newblock \emph{\bibinfo{journal}{Nature}} \bibinfo{pages}{1--8} (\bibinfo{year}{2023}).
\newblock \urlprefix\url{https://www.nature.com/articles/s41586-023-06927-3}.

\bibitem{Loss1998}
\bibinfo{author}{Loss, D.} \& \bibinfo{author}{DiVincenzo, D.~P.}
\newblock \bibinfo{title}{{Quantum computation with quantum dots}}.
\newblock \emph{\bibinfo{journal}{Physical Review A - Atomic, Molecular, and Optical Physics}} \textbf{\bibinfo{volume}{57}}, \bibinfo{pages}{120--126} (\bibinfo{year}{1998}).
\newblock \urlprefix\url{https://journals.aps.org/pra/abstract/10.1103/PhysRevA.57.120}.

\bibitem{Xue2022}
\bibinfo{author}{Xue, X.} \emph{et~al.}
\newblock \bibinfo{title}{{Quantum logic with spin qubits crossing the surface code threshold}}.
\newblock \emph{\bibinfo{journal}{Nature}} \textbf{\bibinfo{volume}{601}}, \bibinfo{pages}{343--347} (\bibinfo{year}{2022}).
\newblock \urlprefix\url{https://www.nature.com/articles/s41586-021-04273-w}.

\bibitem{Weinstein2023}
\bibinfo{author}{Weinstein, A.~J.} \emph{et~al.}
\newblock \bibinfo{title}{{Universal logic with encoded spin qubits in silicon}}.
\newblock \emph{\bibinfo{journal}{Nature}} \textbf{\bibinfo{volume}{615}}, \bibinfo{pages}{817--822} (\bibinfo{year}{2023}).
\newblock \urlprefix\url{https://www.nature.com/articles/s41586-023-05777-3}.

\bibitem{Burkard2023}
\bibinfo{author}{Burkard, G.}, \bibinfo{author}{Ladd, T.~D.}, \bibinfo{author}{Pan, A.}, \bibinfo{author}{Nichol, J.~M.} \& \bibinfo{author}{Petta, J.~R.}
\newblock \bibinfo{title}{{Semiconductor spin qubits}}.
\newblock \emph{\bibinfo{journal}{Reviews of Modern Physics}} \textbf{\bibinfo{volume}{95}}, \bibinfo{pages}{025003} (\bibinfo{year}{2023}).
\newblock \urlprefix\url{https://journals.aps.org/rmp/abstract/10.1103/RevModPhys.95.025003}.

\bibitem{Knill2001}
\bibinfo{author}{Knill, E.}, \bibinfo{author}{Laflamme, R.} \& \bibinfo{author}{Milburn, G.~J.}
\newblock \bibinfo{title}{{A scheme for efficient quantum computation with linear optics}}.
\newblock \emph{\bibinfo{journal}{Nature}} \textbf{\bibinfo{volume}{409}}, \bibinfo{pages}{46--52} (\bibinfo{year}{2001}).
\newblock \urlprefix\url{https://www.nature.com/articles/35051009}.

\bibitem{Browne2005}
\bibinfo{author}{Browne, D.~E.} \& \bibinfo{author}{Rudolph, T.}
\newblock \bibinfo{title}{{Resource-efficient linear optical quantum computation}}.
\newblock \emph{\bibinfo{journal}{Physical Review Letters}} \textbf{\bibinfo{volume}{95}}, \bibinfo{pages}{010501} (\bibinfo{year}{2005}).
\newblock \urlprefix\url{https://journals.aps.org/prl/abstract/10.1103/PhysRevLett.95.010501}.

\bibitem{OBrien2007}
\bibinfo{author}{O'Brien, J.~L.}
\newblock \bibinfo{title}{{Optical quantum computing}}.
\newblock \emph{\bibinfo{journal}{Science}} \textbf{\bibinfo{volume}{318}}, \bibinfo{pages}{1567--1570} (\bibinfo{year}{2007}).
\newblock \urlprefix\url{https://www.science.org/doi/10.1126/science.1142892}.

\bibitem{Shi2022}
\bibinfo{author}{Shi, S.} \emph{et~al.}
\newblock \bibinfo{title}{{High-fidelity photonic quantum logic gate based on near-optimal Rydberg single-photon source}}.
\newblock \emph{\bibinfo{journal}{Nature Communications}} \textbf{\bibinfo{volume}{13}}, \bibinfo{pages}{1--6} (\bibinfo{year}{2022}).
\newblock \urlprefix\url{https://www.nature.com/articles/s41467-022-32083-9}.

\bibitem{Nakamura1999}
\bibinfo{author}{Nakamura, Y.}, \bibinfo{author}{Pashkin, Y.~A.} \& \bibinfo{author}{Tsai, J.~S.}
\newblock \bibinfo{title}{{Coherent control of macroscopic quantum states in a single-Cooper-pair box}}.
\newblock \emph{\bibinfo{journal}{Nature}} \textbf{\bibinfo{volume}{398}}, \bibinfo{pages}{786--788} (\bibinfo{year}{1999}).
\newblock \urlprefix\url{https://www.nature.com/articles/19718}.

\bibitem{Koch2007}
\bibinfo{author}{Koch, J.} \emph{et~al.}
\newblock \bibinfo{title}{{Charge-insensitive qubit design derived from the Cooper pair box}}.
\newblock \emph{\bibinfo{journal}{Physical Review A - Atomic, Molecular, and Optical Physics}} \textbf{\bibinfo{volume}{76}}, \bibinfo{pages}{042319} (\bibinfo{year}{2007}).
\newblock \urlprefix\url{https://journals.aps.org/pra/abstract/10.1103/PhysRevA.76.042319}.

\bibitem{Devoret2005}
\bibinfo{author}{Devoret, M.~H.} \& \bibinfo{author}{Martinis, J.~M.}
\newblock \bibinfo{title}{{Implementing qubits with superconducting integrated circuits}}.
\newblock \emph{\bibinfo{journal}{Experimental Aspects of Quantum Computing}} \textbf{\bibinfo{volume}{3}}, \bibinfo{pages}{163--203} (\bibinfo{year}{2005}).
\newblock \urlprefix\url{https://link.springer.com/article/10.1007/s11128-004-3101-5}.

\bibitem{Kockum2019}
\bibinfo{author}{Kockum, A.~F.} \& \bibinfo{author}{Nori, F.}
\newblock \bibinfo{title}{{Quantum Bits with Josephson Junctions}}.
\newblock \emph{\bibinfo{journal}{Springer Series in Materials Science}} \textbf{\bibinfo{volume}{286}}, \bibinfo{pages}{703--741} (\bibinfo{year}{2019}).
\newblock \urlprefix\url{https://link.springer.com/chapter/10.1007/978-3-030-20726-7_17}.

\bibitem{Mooij1999}
\bibinfo{author}{Mooij, J.~E.} \emph{et~al.}
\newblock \bibinfo{title}{{Josephson persistent-current qubit}}.
\newblock \emph{\bibinfo{journal}{Science}} \textbf{\bibinfo{volume}{285}}, \bibinfo{pages}{1036--1039} (\bibinfo{year}{1999}).
\newblock \urlprefix\url{https://www.science.org/doi/10.1126/science.285.5430.1036}.

\bibitem{Friedman2000}
\bibinfo{author}{Friedman, J.~R.}, \bibinfo{author}{Patel, V.}, \bibinfo{author}{Chen, W.}, \bibinfo{author}{Tolpygo, S.~K.} \& \bibinfo{author}{Lukens, J.~E.}
\newblock \bibinfo{title}{{Quantum superposition of distinct macroscopic states}}.
\newblock \emph{\bibinfo{journal}{Nature}} \textbf{\bibinfo{volume}{406}}, \bibinfo{pages}{43--46} (\bibinfo{year}{2000}).
\newblock \urlprefix\url{https://www.nature.com/articles/35017505}.

\bibitem{Vion2002}
\bibinfo{author}{Vion, D.} \emph{et~al.}
\newblock \bibinfo{title}{{Manipulating the quantum state of an electrical circuit}}.
\newblock \emph{\bibinfo{journal}{Science}} \textbf{\bibinfo{volume}{296}}, \bibinfo{pages}{886--889} (\bibinfo{year}{2002}).
\newblock \urlprefix\url{https://www.science.org/doi/10.1126/science.1069372}.

\bibitem{Han2001}
\bibinfo{author}{Han, S.}, \bibinfo{author}{Yu, Y.}, \bibinfo{author}{Chu, X.}, \bibinfo{author}{Chu, S.~I.} \& \bibinfo{author}{Wang, Z.}
\newblock \bibinfo{title}{{Time-resolved measurement of dissipation-induced decoherence in a Josephson junction}}.
\newblock \emph{\bibinfo{journal}{Science}} \textbf{\bibinfo{volume}{293}}, \bibinfo{pages}{1457--1459} (\bibinfo{year}{2001}).
\newblock \urlprefix\url{https://www.science.org/doi/10.1126/science.1062266}.

\bibitem{Simmonds2004}
\bibinfo{author}{Simmonds, R.~W.} \emph{et~al.}
\newblock \bibinfo{title}{{Decoherence in josephson phase qubits from junction resonators}}.
\newblock \emph{\bibinfo{journal}{Physical Review Letters}} \textbf{\bibinfo{volume}{93}}, \bibinfo{pages}{077003} (\bibinfo{year}{2004}).
\newblock \urlprefix\url{https://journals.aps.org/prl/abstract/10.1103/PhysRevLett.93.077003}.

\bibitem{Wang2022}
\bibinfo{author}{Wang, C.} \emph{et~al.}
\newblock \bibinfo{title}{{Towards practical quantum computers: transmon qubit with a lifetime approaching 0.5 milliseconds}}.
\newblock \emph{\bibinfo{journal}{npj Quantum Information}} \textbf{\bibinfo{volume}{8}}, \bibinfo{pages}{1--6} (\bibinfo{year}{2022}).
\newblock \urlprefix\url{https://www.nature.com/articles/s41534-021-00510-2}.

\bibitem{Verjauw2022}
\bibinfo{author}{Verjauw, J.} \emph{et~al.}
\newblock \bibinfo{title}{{Path toward manufacturable superconducting qubits with relaxation times exceeding 0.1 ms}}.
\newblock \emph{\bibinfo{journal}{npj Quantum Information}} \textbf{\bibinfo{volume}{8}}, \bibinfo{pages}{1--7} (\bibinfo{year}{2022}).
\newblock \urlprefix\url{https://www.nature.com/articles/s41534-022-00600-9}.

\bibitem{Place2021}
\bibinfo{author}{Place, A.~P.} \emph{et~al.}
\newblock \bibinfo{title}{{New material platform for superconducting transmon qubits with coherence times exceeding 0.3 milliseconds}}.
\newblock \emph{\bibinfo{journal}{Nature Communications}} \textbf{\bibinfo{volume}{12}}, \bibinfo{pages}{1--6} (\bibinfo{year}{2021}).
\newblock \urlprefix\url{https://www.nature.com/articles/s41467-021-22030-5}.

\bibitem{Somoroff2023}
\bibinfo{author}{Somoroff, A.} \emph{et~al.}
\newblock \bibinfo{title}{{Millisecond Coherence in a Superconducting Qubit}}.
\newblock \emph{\bibinfo{journal}{Physical Review Letters}} \textbf{\bibinfo{volume}{130}}, \bibinfo{pages}{267001} (\bibinfo{year}{2023}).
\newblock \urlprefix\url{https://journals.aps.org/prl/abstract/10.1103/PhysRevLett.130.267001}.

\bibitem{Werninghaus2021}
\bibinfo{author}{Werninghaus, M.} \emph{et~al.}
\newblock \bibinfo{title}{{Leakage reduction in fast superconducting qubit gates via optimal control}}.
\newblock \emph{\bibinfo{journal}{npj Quantum Information}} \textbf{\bibinfo{volume}{7}}, \bibinfo{pages}{1--6} (\bibinfo{year}{2021}).
\newblock \urlprefix\url{https://www.nature.com/articles/s41534-020-00346-2}.

\bibitem{Ding2023}
\bibinfo{author}{Ding, L.} \emph{et~al.}
\newblock \bibinfo{title}{{High-Fidelity, Frequency-Flexible Two-Qubit Fluxonium Gates with a Transmon Coupler}}.
\newblock \emph{\bibinfo{journal}{Physical Review X}} \textbf{\bibinfo{volume}{13}}, \bibinfo{pages}{031035} (\bibinfo{year}{2023}).
\newblock \urlprefix\url{https://journals.aps.org/prx/abstract/10.1103/PhysRevX.13.031035}.

\bibitem{Martinis2005}
\bibinfo{author}{Martinis, J.~M.} \emph{et~al.}
\newblock \bibinfo{title}{{Decoherence in Josephson qubits from dielectric Loss}}.
\newblock \emph{\bibinfo{journal}{Physical Review Letters}} \textbf{\bibinfo{volume}{95}}, \bibinfo{pages}{210503} (\bibinfo{year}{2005}).
\newblock \urlprefix\url{https://journals.aps.org/prl/abstract/10.1103/PhysRevLett.95.210503}.

\bibitem{Gao2008}
\bibinfo{author}{Gao, J.} \emph{et~al.}
\newblock \bibinfo{title}{{A semiempirical model for two-level system noise in superconducting microresonators}}.
\newblock \emph{\bibinfo{journal}{Applied Physics Letters}} \textbf{\bibinfo{volume}{92}}, \bibinfo{pages}{212504} (\bibinfo{year}{2008}).
\newblock \urlprefix\url{/aip/apl/article/92/21/212504/851718/A-semiempirical-model-for-two-level-system-noise}.

\bibitem{Lindstrom2009}
\bibinfo{author}{Lindstr{\"{o}}m, T.}, \bibinfo{author}{Healey, J.~E.}, \bibinfo{author}{Colclough, M.~S.}, \bibinfo{author}{Muirhead, C.~M.} \& \bibinfo{author}{Tzalenchuk, A.~Y.}
\newblock \bibinfo{title}{{Properties of superconducting planar resonators at millikelvin temperatures}}.
\newblock \emph{\bibinfo{journal}{Physical Review B - Condensed Matter and Materials Physics}} \textbf{\bibinfo{volume}{80}}, \bibinfo{pages}{132501} (\bibinfo{year}{2009}).
\newblock \urlprefix\url{https://journals.aps.org/prb/abstract/10.1103/PhysRevB.80.132501}.

\bibitem{Macha2010}
\bibinfo{author}{Macha, P.} \emph{et~al.}
\newblock \bibinfo{title}{{Losses in coplanar waveguide resonators at millikelvin temperatures}}.
\newblock \emph{\bibinfo{journal}{Applied Physics Letters}} \textbf{\bibinfo{volume}{96}}, \bibinfo{pages}{62503} (\bibinfo{year}{2010}).
\newblock \urlprefix\url{/aip/apl/article/96/6/062503/167025/Losses-in-coplanar-waveguide-resonators-at}.

\bibitem{Pappas2011}
\bibinfo{author}{Pappas, D.~P.}, \bibinfo{author}{Vissers, M.~R.}, \bibinfo{author}{Wisbey, D.~S.}, \bibinfo{author}{Kline, J.~S.} \& \bibinfo{author}{Gao, J.}
\newblock \bibinfo{title}{{Two level system loss in superconducting microwave resonators}}.
\newblock \emph{\bibinfo{journal}{IEEE Transactions on Applied Superconductivity}} \textbf{\bibinfo{volume}{21}}, \bibinfo{pages}{871--874} (\bibinfo{year}{2011}).

\bibitem{Neill2013}
\bibinfo{author}{Neill, C.} \emph{et~al.}
\newblock \bibinfo{title}{{Fluctuations from edge defects in superconducting resonators}}.
\newblock \emph{\bibinfo{journal}{Applied Physics Letters}} \textbf{\bibinfo{volume}{103}}, \bibinfo{pages}{72601} (\bibinfo{year}{2013}).
\newblock \urlprefix\url{/aip/apl/article/103/7/072601/150361/Fluctuations-from-edge-defects-in-superconducting}.

\bibitem{Muller2015}
\bibinfo{author}{M{\"{u}}ller, C.}, \bibinfo{author}{Lisenfeld, J.}, \bibinfo{author}{Shnirman, A.} \& \bibinfo{author}{Poletto, S.}
\newblock \bibinfo{title}{{Interacting two-level defects as sources of fluctuating high-frequency noise in superconducting circuits}}.
\newblock \emph{\bibinfo{journal}{Physical Review B - Condensed Matter and Materials Physics}} \textbf{\bibinfo{volume}{92}}, \bibinfo{pages}{035442} (\bibinfo{year}{2015}).
\newblock \urlprefix\url{https://journals.aps.org/prb/abstract/10.1103/PhysRevB.92.035442}.

\bibitem{Wang2015}
\bibinfo{author}{Wang, C.} \emph{et~al.}
\newblock \bibinfo{title}{{Surface participation and dielectric loss in superconducting qubits}}.
\newblock \emph{\bibinfo{journal}{Applied Physics Letters}} \textbf{\bibinfo{volume}{107}}, \bibinfo{pages}{162601} (\bibinfo{year}{2015}).
\newblock \urlprefix\url{/aip/apl/article/107/16/162601/593971/Surface-participation-and-dielectric-loss-in}.

\bibitem{Lutchyn2006}
\bibinfo{author}{Lutchyn, R.~M.}, \bibinfo{author}{Glazman, L.~I.} \& \bibinfo{author}{Larkin, A.~I.}
\newblock \bibinfo{title}{{Kinetics of the superconducting charge qubit in the presence of a quasiparticle}}.
\newblock \emph{\bibinfo{journal}{Physical Review B - Condensed Matter and Materials Physics}} \textbf{\bibinfo{volume}{74}}, \bibinfo{pages}{064515} (\bibinfo{year}{2006}).
\newblock \urlprefix\url{https://journals.aps.org/prb/abstract/10.1103/PhysRevB.74.064515}.

\bibitem{Paik2011}
\bibinfo{author}{Paik, H.} \emph{et~al.}
\newblock \bibinfo{title}{{Observation of high coherence in Josephson junction qubits measured in a three-dimensional circuit QED architecture}}.
\newblock \emph{\bibinfo{journal}{Physical Review Letters}} \textbf{\bibinfo{volume}{107}}, \bibinfo{pages}{240501} (\bibinfo{year}{2011}).
\newblock \urlprefix\url{https://journals.aps.org/prl/abstract/10.1103/PhysRevLett.107.240501}.

\bibitem{Catelani2011}
\bibinfo{author}{Catelani, G.} \emph{et~al.}
\newblock \bibinfo{title}{{Quasiparticle relaxation of superconducting qubits in the presence of flux}}.
\newblock \emph{\bibinfo{journal}{Physical Review Letters}} \textbf{\bibinfo{volume}{106}}, \bibinfo{pages}{077002} (\bibinfo{year}{2011}).
\newblock \urlprefix\url{https://journals.aps.org/prl/abstract/10.1103/PhysRevLett.106.077002}.

\bibitem{Catelani2012}
\bibinfo{author}{Catelani, G.}, \bibinfo{author}{Nigg, S.~E.}, \bibinfo{author}{Girvin, S.~M.}, \bibinfo{author}{Schoelkopf, R.~J.} \& \bibinfo{author}{Glazman, L.~I.}
\newblock \bibinfo{title}{{Decoherence of superconducting qubits caused by quasiparticle tunneling}}.
\newblock \emph{\bibinfo{journal}{Physical Review B - Condensed Matter and Materials Physics}} \textbf{\bibinfo{volume}{86}}, \bibinfo{pages}{184514} (\bibinfo{year}{2012}).
\newblock \urlprefix\url{https://journals.aps.org/prb/abstract/10.1103/PhysRevB.86.184514}.

\bibitem{Paladino2014}
\bibinfo{author}{Paladino, E.}, \bibinfo{author}{Galperin, Y.}, \bibinfo{author}{Falci, G.} \& \bibinfo{author}{Altshuler, B.~L.}
\newblock \bibinfo{title}{{1/ f noise: Implications for solid-state quantum information}}.
\newblock \emph{\bibinfo{journal}{Reviews of Modern Physics}} \textbf{\bibinfo{volume}{86}}, \bibinfo{pages}{361--418} (\bibinfo{year}{2014}).
\newblock \urlprefix\url{https://journals.aps.org/rmp/abstract/10.1103/RevModPhys.86.361}.

\bibitem{Burnett2014}
\bibinfo{author}{Burnett, J.} \emph{et~al.}
\newblock \bibinfo{title}{{Evidence for interacting two-level systems from the 1/f noise of a superconducting resonator}}.
\newblock \emph{\bibinfo{journal}{Nature Communications}} \textbf{\bibinfo{volume}{5}}, \bibinfo{pages}{1--6} (\bibinfo{year}{2014}).
\newblock \urlprefix\url{https://www.nature.com/articles/ncomms5119}.

\bibitem{Pop2014}
\bibinfo{author}{Pop, I.~M.} \emph{et~al.}
\newblock \bibinfo{title}{{Coherent suppression of electromagnetic dissipation due to superconducting quasiparticles}}.
\newblock \emph{\bibinfo{journal}{Nature}} \textbf{\bibinfo{volume}{508}}, \bibinfo{pages}{369--372} (\bibinfo{year}{2014}).
\newblock \urlprefix\url{https://www.nature.com/articles/nature13017}.

\bibitem{Lisenfeld2019}
\bibinfo{author}{Lisenfeld, J.} \emph{et~al.}
\newblock \bibinfo{title}{{Electric field spectroscopy of material defects in transmon qubits}}.
\newblock \emph{\bibinfo{journal}{npj Quantum Information}} \textbf{\bibinfo{volume}{5}}, \bibinfo{pages}{1--6} (\bibinfo{year}{2019}).
\newblock \urlprefix\url{https://www.nature.com/articles/s41534-019-0224-1}.

\bibitem{Yan2016}
\bibinfo{author}{Yan, F.} \emph{et~al.}
\newblock \bibinfo{title}{{The flux qubit revisited to enhance coherence and reproducibility}}.
\newblock \emph{\bibinfo{journal}{Nature Communications}} \textbf{\bibinfo{volume}{7}}, \bibinfo{pages}{1--9} (\bibinfo{year}{2016}).
\newblock \urlprefix\url{https://www.nature.com/articles/ncomms12964}.

\bibitem{Vepsalainen2020}
\bibinfo{author}{Veps{\"{a}}l{\"{a}}inen, A.~P.} \emph{et~al.}
\newblock \bibinfo{title}{{Impact of ionizing radiation on superconducting qubit coherence}}.
\newblock \emph{\bibinfo{journal}{Nature}} \textbf{\bibinfo{volume}{584}}, \bibinfo{pages}{551--556} (\bibinfo{year}{2020}).
\newblock \urlprefix\url{https://www.nature.com/articles/s41586-020-2619-8}.

\bibitem{Murray2021}
\bibinfo{author}{Murray, C.~E.}
\newblock \bibinfo{title}{{Material matters in superconducting qubits}}.
\newblock \emph{\bibinfo{journal}{Materials Science and Engineering R: Reports}} \textbf{\bibinfo{volume}{146}}, \bibinfo{pages}{100646} (\bibinfo{year}{2021}).

\bibitem{Siddiqi2021}
\bibinfo{author}{Siddiqi, I.}
\newblock \bibinfo{title}{{Engineering high-coherence superconducting qubits}}.
\newblock \emph{\bibinfo{journal}{Nature Reviews Materials}} \textbf{\bibinfo{volume}{6}}, \bibinfo{pages}{875--891} (\bibinfo{year}{2021}).
\newblock \urlprefix\url{https://www.nature.com/articles/s41578-021-00370-4}.

\bibitem{Carroll2022}
\bibinfo{author}{Carroll, M.}, \bibinfo{author}{Rosenblatt, S.}, \bibinfo{author}{Jurcevic, P.}, \bibinfo{author}{Lauer, I.} \& \bibinfo{author}{Kandala, A.}
\newblock \bibinfo{title}{{Dynamics of superconducting qubit relaxation times}}.
\newblock \emph{\bibinfo{journal}{npj Quantum Information}} \textbf{\bibinfo{volume}{8}}, \bibinfo{pages}{1--7} (\bibinfo{year}{2022}).
\newblock \urlprefix\url{https://www.nature.com/articles/s41534-022-00643-y}.

\bibitem{Wang2009}
\bibinfo{author}{Wang, H.} \emph{et~al.}
\newblock \bibinfo{title}{{Improving the coherence time of superconducting coplanar resonators}}.
\newblock \emph{\bibinfo{journal}{Applied Physics Letters}} \textbf{\bibinfo{volume}{95}}, \bibinfo{pages}{233508} (\bibinfo{year}{2009}).
\newblock \urlprefix\url{/aip/apl/article/95/23/233508/120944/Improving-the-coherence-time-of-superconducting}.

\bibitem{Megrant2012}
\bibinfo{author}{Megrant, A.} \emph{et~al.}
\newblock \bibinfo{title}{{Planar superconducting resonators with internal quality factors above one million}}.
\newblock \emph{\bibinfo{journal}{Applied Physics Letters}} \textbf{\bibinfo{volume}{100}}, \bibinfo{pages}{113510} (\bibinfo{year}{2012}).
\newblock \urlprefix\url{/aip/apl/article/100/11/113510/126200/Planar-superconducting-resonators-with-internal}.

\bibitem{Chang2013}
\bibinfo{author}{Chang, J.~B.} \emph{et~al.}
\newblock \bibinfo{title}{{Improved superconducting qubit coherence using titanium nitride}}.
\newblock \emph{\bibinfo{journal}{Applied Physics Letters}} \textbf{\bibinfo{volume}{103}}, \bibinfo{pages}{12602} (\bibinfo{year}{2013}).
\newblock \urlprefix\url{/aip/apl/article/103/1/012602/235870/Improved-superconducting-qubit-coherence-using}.

\bibitem{Stern2014}
\bibinfo{author}{Stern, M.} \emph{et~al.}
\newblock \bibinfo{title}{{Flux qubits with long coherence times for hybrid quantum circuits}}.
\newblock \emph{\bibinfo{journal}{Physical Review Letters}} \textbf{\bibinfo{volume}{113}}, \bibinfo{pages}{123601} (\bibinfo{year}{2014}).
\newblock \urlprefix\url{https://journals.aps.org/prl/abstract/10.1103/PhysRevLett.113.123601}.

\bibitem{Bruno2015}
\bibinfo{author}{Bruno, A.} \emph{et~al.}
\newblock \bibinfo{title}{{Reducing intrinsic loss in superconducting resonators by surface treatment and deep etching of silicon substrates}}.
\newblock \emph{\bibinfo{journal}{Applied Physics Letters}} \textbf{\bibinfo{volume}{106}}, \bibinfo{pages}{182601} (\bibinfo{year}{2015}).
\newblock \urlprefix\url{/aip/apl/article/106/18/182601/27784/Reducing-intrinsic-loss-in-superconducting}.

\bibitem{Deng2023}
\bibinfo{author}{Deng, H.} \emph{et~al.}
\newblock \bibinfo{title}{{Titanium Nitride Film on Sapphire Substrate with Low Dielectric Loss for Superconducting Qubits}}.
\newblock \emph{\bibinfo{journal}{Physical Review Applied}} \textbf{\bibinfo{volume}{19}}, \bibinfo{pages}{024013} (\bibinfo{year}{2023}).
\newblock \urlprefix\url{https://journals.aps.org/prapplied/abstract/10.1103/PhysRevApplied.19.024013}.

\bibitem{He2022}
\bibinfo{author}{He, H.}, \bibinfo{author}{Wang, W.}, \bibinfo{author}{Liu, F.}, \bibinfo{author}{Yuan, B.} \& \bibinfo{author}{Shan, Z.}
\newblock \bibinfo{title}{{Suppressing the Dielectric Loss in Superconducting Qubits through Useful Geometry Design}}.
\newblock \emph{\bibinfo{journal}{Entropy}} \textbf{\bibinfo{volume}{24}}, \bibinfo{pages}{952} (\bibinfo{year}{2022}).
\newblock \urlprefix\url{https://www.mdpi.com/1099-4300/24/7/952/htm https://www.mdpi.com/1099-4300/24/7/952}.

\bibitem{Martinis2022}
\bibinfo{author}{Martinis, J.~M.}
\newblock \bibinfo{title}{{Surface loss calculations and design of a superconducting transmon qubit with tapered wiring}}.
\newblock \emph{\bibinfo{journal}{npj Quantum Information}} \textbf{\bibinfo{volume}{8}}, \bibinfo{pages}{1--12} (\bibinfo{year}{2022}).
\newblock \urlprefix\url{https://www.nature.com/articles/s41534-022-00530-6}.

\bibitem{Krinner}
\bibinfo{author}{Krinner, S.} \emph{et~al.}
\newblock \bibinfo{title}{{Engineering cryogenic setups for 100-qubit scale superconducting circuit systems}}.
\newblock \emph{\bibinfo{journal}{EPJ Quantum Technology}} \textbf{\bibinfo{volume}{6}} (\bibinfo{year}{2019}).
\newblock \urlprefix\url{https://doi.org/10.1140/epjqt/s40507-019-0072-0}.

\bibitem{Kono2020}
\bibinfo{author}{Kono, S.} \emph{et~al.}
\newblock \bibinfo{title}{{Breaking the trade-off between fast control and long lifetime of a superconducting qubit}}.
\newblock \emph{\bibinfo{journal}{Nature Communications}} \textbf{\bibinfo{volume}{11}}, \bibinfo{pages}{1--6} (\bibinfo{year}{2020}).
\newblock \urlprefix\url{https://www.nature.com/articles/s41467-020-17511-y}.

\bibitem{Xia2023}
\bibinfo{author}{Xia, M.} \emph{et~al.}
\newblock \bibinfo{title}{{Fast superconducting qubit control with sub-harmonic drives}}  (\bibinfo{year}{2023}).
\newblock \urlprefix\url{http://arxiv.org/abs/2306.10162}.

\bibitem{FriskKockum2014}
\bibinfo{author}{Frisk~Kockum, A.}, \bibinfo{author}{Delsing, P.} \& \bibinfo{author}{Johansson, G.}
\newblock \bibinfo{title}{{Designing frequency-dependent relaxation rates and Lamb shifts for a giant artificial atom}}.
\newblock \emph{\bibinfo{journal}{Physical Review A - Atomic, Molecular, and Optical Physics}} \textbf{\bibinfo{volume}{90}}, \bibinfo{pages}{013837} (\bibinfo{year}{2014}).
\newblock \urlprefix\url{https://journals.aps.org/pra/abstract/10.1103/PhysRevA.90.013837}.

\bibitem{Kannan2020}
\bibinfo{author}{Kannan, B.} \emph{et~al.}
\newblock \bibinfo{title}{{Waveguide quantum electrodynamics with superconducting artificial giant atoms}}.
\newblock \emph{\bibinfo{journal}{Nature}} \textbf{\bibinfo{volume}{583}}, \bibinfo{pages}{775--779} (\bibinfo{year}{2020}).
\newblock \urlprefix\url{https://www.nature.com/articles/s41586-020-2529-9}.

\bibitem{Barends2013}
\bibinfo{author}{Barends, R.} \emph{et~al.}
\newblock \bibinfo{title}{{Coherent josephson qubit suitable for scalable quantum integrated circuits}}.
\newblock \emph{\bibinfo{journal}{Physical Review Letters}} \textbf{\bibinfo{volume}{111}}, \bibinfo{pages}{080502} (\bibinfo{year}{2013}).
\newblock \urlprefix\url{https://journals.aps.org/prl/abstract/10.1103/PhysRevLett.111.080502}.

\bibitem{Jeffrey2014}
\bibinfo{author}{Jeffrey, E.} \emph{et~al.}
\newblock \bibinfo{title}{{Fast accurate state measurement with superconducting qubits}}.
\newblock \emph{\bibinfo{journal}{Physical Review Letters}} \textbf{\bibinfo{volume}{112}}, \bibinfo{pages}{190504} (\bibinfo{year}{2014}).
\newblock \urlprefix\url{https://journals.aps.org/prl/abstract/10.1103/PhysRevLett.112.190504}.

\bibitem{Esteve1986}
\bibinfo{author}{Esteve, D.}, \bibinfo{author}{Devoret, M.~H.} \& \bibinfo{author}{Martinis, J.~M.}
\newblock \bibinfo{title}{{Effect of an arbitrary dissipative circuit on the quantum energy levels and tunneling of a Josephson junction}}.
\newblock \emph{\bibinfo{journal}{Physical Review B}} \textbf{\bibinfo{volume}{34}}, \bibinfo{pages}{158--163} (\bibinfo{year}{1986}).
\newblock \urlprefix\url{https://journals.aps.org/prb/abstract/10.1103/PhysRevB.34.158}.

\bibitem{Houck2008a}
\bibinfo{author}{Houck, A.~A.} \emph{et~al.}
\newblock \bibinfo{title}{{Controlling the spontaneous emission of a superconducting transmon qubit}}.
\newblock \emph{\bibinfo{journal}{Physical Review Letters}} \textbf{\bibinfo{volume}{101}}, \bibinfo{pages}{080502} (\bibinfo{year}{2008}).
\newblock \urlprefix\url{https://journals.aps.org/prl/abstract/10.1103/PhysRevLett.101.080502}.

\bibitem{Blais2021a}
\bibinfo{author}{Blais, A.}, \bibinfo{author}{Grimsmo, A.~L.}, \bibinfo{author}{Girvin, S.~M.} \& \bibinfo{author}{Wallraff, A.}
\newblock \bibinfo{title}{{Circuit quantum electrodynamics}}.
\newblock \emph{\bibinfo{journal}{Reviews of Modern Physics}} \textbf{\bibinfo{volume}{93}}, \bibinfo{pages}{025005} (\bibinfo{year}{2021}).

\bibitem{sah_2024_11234843}
\bibinfo{author}{Sah, A.}, \bibinfo{author}{Kundu, S.}, \bibinfo{author}{Suominen, H.}, \bibinfo{author}{Chen, Q.} \& \bibinfo{author}{M{\"{o}}tt{\"{o}}nen, M.}
\newblock \bibinfo{title}{{Data and codes for "Decay-protected superconducting qubit with fast control enabled by integrated on-chip filters"}} (\bibinfo{year}{2024}).
\newblock \urlprefix\url{https://doi.org/10.5281/zenodo.11234843}.

\bibitem{Fedorov2019}
\bibinfo{author}{Fedorov, G.~P.} \& \bibinfo{author}{Ustinov, A.~V.}
\newblock \bibinfo{title}{{Automated analysis of single-tone spectroscopic data for cQED systems}}.
\newblock \emph{\bibinfo{journal}{Quantum Science and Technology}} \textbf{\bibinfo{volume}{4}}, \bibinfo{pages}{045009} (\bibinfo{year}{2019}).
\newblock \urlprefix\url{https://iopscience.iop.org/article/10.1088/2058-9565/ab478b https://iopscience.iop.org/article/10.1088/2058-9565/ab478b/meta}.

\bibitem{Zhang2021}
\bibinfo{author}{Zhang, H.} \emph{et~al.}
\newblock \bibinfo{title}{{Universal Fast-Flux Control of a Coherent, Low-Frequency Qubit}}.
\newblock \emph{\bibinfo{journal}{Physical Review X}} \textbf{\bibinfo{volume}{11}}, \bibinfo{pages}{011010} (\bibinfo{year}{2021}).
\newblock \urlprefix\url{https://journals.aps.org/prx/abstract/10.1103/PhysRevX.11.011010}.

\bibitem{Hyyppa2022}
\bibinfo{author}{Hyypp{\"{a}}, E.} \emph{et~al.}
\newblock \bibinfo{title}{{Unimon qubit}}.
\newblock \emph{\bibinfo{journal}{Nature Communications}} \textbf{\bibinfo{volume}{13}}, \bibinfo{pages}{1--14} (\bibinfo{year}{2022}).
\newblock \urlprefix\url{https://www.nature.com/articles/s41467-022-34614-w}.

\end{thebibliography}

\section*{Acknowledgments}
We express our gratitude to the KAUTE Foundation for their support through the Ph.D. grant awarded to the first author. We acknowledge the European Research Council under the Advanced Grant no. 101053801 (ConceptQ), Business Finland under the Quantum Technologies Industrial project (Grant no. 2118781) and Academy of Finland under its Centre of Excellence Quantum Technology Finland (Grant no. 352925) and through the Finnish Quantum Flagship. Special thanks are extended to Sergei Lemziakov, Dmitrii Lvov, and Joonas Peltonen for their invaluable assistance with the Otanano cryogenic facility. The authors acknowledge Giacomo Catto, Timm Mörstedt, and Priyank Singh for useful discussions related to the fabrication and Jukka-Pekka Kaikkonen from VTT Technical Research for providing us with the unpatterned Niobium-on-Silicon wafer. Furthermore, we are thankful for the enriching general discussions on the subject matter with Michael Hatridge, Florian Blanchet, Arman Alizadeh, Wallace Santos Teixeira, Vasilii Vadimov, and Johannes Heinsoo. We thank Eric Hyyppä for encouraging discussions and insightful comments on the manuscript.

\section*{Authors contributions}
A.S. and M.M. conceived the experiment. A.S. and S.K. designed and carried out the electromagnetic simulations of the device. H.S. and A.S. wrote the measurement codes for the experiment. Q.C. helped with the theoretical discussions on subharmonic drive. A.S. fabricated the device, conducted the experiments, and analyzed the results with feedback from S.K. A.S., S.K., and M.M. wrote the manuscript with comments from all the authors.

\section*{Competing interests}
M.M. declares that he is a Co-Founder and Shareholder of IQM Finland Oy. All other authors declare no competing interests.

\clearpage 
\appendix 
\pagenumbering{roman} 
\setcounter{page}{1} 

\twocolumn[\section*{Supplementary Information for "Decay-protected superconducting qubit with fast control enabled by integrated on-chip filters"}] 
\renewcommand{\thefigure}{\textbf{S\arabic{figure}}}
\renewcommand{\thetable}{\textbf{S\arabic{table}}}
\renewcommand{\figurename}{\textbf{Supplementary Fig.}} 
\setcounter{figure}{0} 
\renewcommand{\tablename}{\textbf{Supplementary Table}} 
\setcounter{table}{0} 

\section*{Supplementary Note S1: Design and simulation of the on-chip filters}

To facilitate the design and the implementation of the on-chip filters, we establish a simulation procedure to obtain the Rabi frequency and the $T_\mathrm{1, ext}$ of the qubit resulting from the coupling to the on-chip filters. To this end, we estimate the quantum properties of superconducting circuits from classical electromagnetic simulations of lumped and distributed-element circuits~\cite{Esteve1986, Houck2008a, Blais2021a}, see Supplementary Fig.~\ref{fig:comparision_Rabi_T1}a-c. By simulating the input admittance ($Y_\mathrm{in}$) in parallel with the qubit from qubit--environment coupling, one obtains the external coupling rate using \ensuremath{\gamma_\mathrm{q}(\omega) = \mathrm{Re}[Y_\mathrm{in}(\omega)]/C_\mathrm{q}}, where $C_\mathrm{q}$ is the shunt capacitance of the qubit to the ground. Here, the environment refers to any element that couples to the qubit such as standard drive lines, on-chip filters, readout resonators, and flux lines. However, care needs to be taken when addressing the contribution from the individual coupling elements, for example, by tuning out the other coupled channels by either grounding them or removing them completely from the simulation geometry.

\begin{figure*}[ht]
\includegraphics[width=\linewidth]{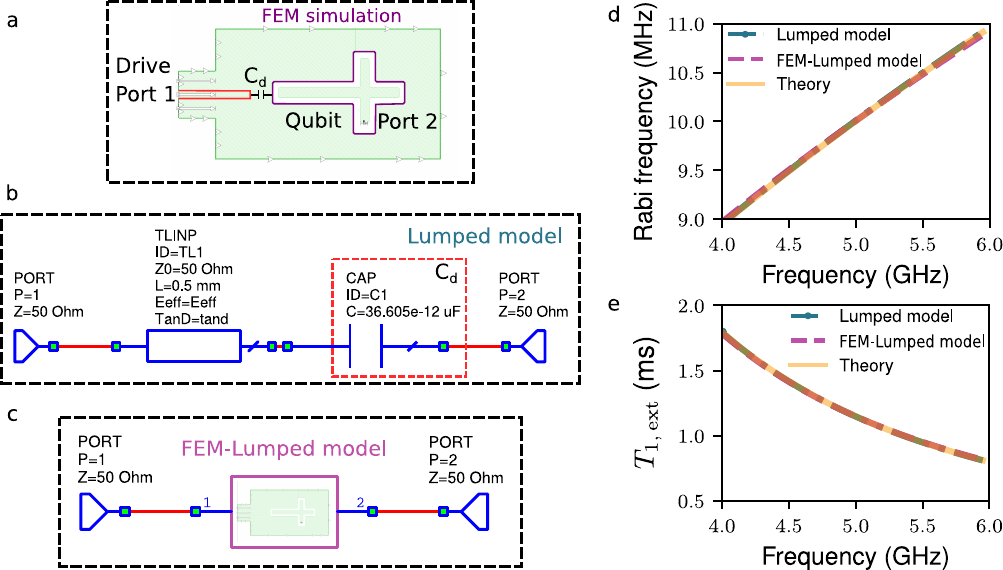}
\caption{\textbf{Comparison different models for the Rabi frequency and the energy decay time of a weakly coupled qubit to a standard drive.} \textbf{a} Simplified model of a transmon qubit coupled to a coplanar-waveguide (CPW) transmission line rendered in Sonnet. A box-wall port 1 is used at the drive port, and a co-calibrated lumped port 2 is placed on the qubit island to investigate scattering, $S_\mathrm{21}$, and input admittance, $Y_\mathrm{in}$, parameters of the qubit. Note that $Y_\mathrm{in}$ is calculated when the other ports are terminated using the impedances set by the port terminations, in this case, to 50~\ensuremath{\Omega}. An estimate of the qubit-drive coupling capacitance \ensuremath{C_\mathrm{d}} is obtained from the finite-element method (FEM) simulation carried out in Sonnet. \textbf{b} Equivalent lumped model of the qubit-drive system built in Microwave Office AWR design environment. A transmission line element is used to model a CPW drive line with a characteristic impedance $Z_\mathrm{0}$ of 50~\ensuremath{\Omega} and relative effective dielectric constant $E_\mathrm{eff}$ of 6.45 obtained using \ensuremath{E_\mathrm{eff} = (\epsilon_\mathrm{Si} + \epsilon_\mathrm{air})/2}, where \ensuremath{\epsilon_\mathrm{Si} = 11.9} and \ensuremath{\epsilon_\mathrm{air} = 1} are the relative dielectric permittivity of a silicon substrate and air, respectively. The coupling capacitance \ensuremath{C_\mathrm{d}} is taken from the Sonnet simulation. \textbf{c} Combination of a FEM simulation and a lumped model is set up inside AWR, allowing us to simulate complex and large geometries while avoiding the time and memory constraints usually incurred in FEM simulations of such geometries. From Sonnet simulation, we obtain a S-parameter file in touchstone format and export it to AWR, where FEM-lumped-model simulation yields $Y_\mathrm{in}$ in parallel with the qubit due to the coupling to the drive. \textbf{d, e} Comparison of the estimated (\textbf{d}) Rabi frequency and (\textbf{e}) Relaxation time, $T_\mathrm{1, ext}$, of the qubit from lumped, FEM-lumped, and analytical approaches.
}
\label{fig:comparision_Rabi_T1}
\end{figure*}

Once the external coupling rate is known, the Rabi frequency and the $T_\mathrm{1, ext}$ can be estimated using \ensuremath{\Omega_\mathrm{R}(\omega) = 2\sqrt{\gamma_\mathrm{q}(\omega)}\beta} and \ensuremath{T_\mathrm{1, ext}(\omega) = 1/\gamma_\mathrm{q}(\omega)}, respectively, as discussed in the Methods of the main text. Using this approach, we compare the Rabi frequency and the $T_\mathrm{1, ext}$ estimates obtained from FEM and lumped-element simulations to the theoretical model for a qubit weakly coupled to a standard transmission line, as shown in Supplementary Fig.~\ref{fig:comparision_Rabi_T1}d,e. We observe a very good correspondence between the simulated and the theoretical results, confirming the validity of the established approach. 

In Supplementary Fig.~\ref{fig:simulation_half_wave_filter}, we describe the approach for full-circuit simulation for the qubit Q3 coupled to the \ensuremath{\lambda/2} filter, the readout resonator, and the flux line. Supplementary Figure~\ref{fig:simulation_half_wave_filter}a shows a simplified model of a large simulation geometry, depicted in Supplementary Fig.~\ref{fig:simulation_half_wave_filter}b, by trimming out the meanders of the \ensuremath{\lambda/2} filter and adding co-calibrated lumped ports that allow for building the full-circuit scheme with a lumped-element network simulator such as Microwave office AWR. After building the full-circuit scheme in AWR, we simulate the input admittance in parallel with the qubit from the port of interest, as shown in Supplementary Fig.~\ref{fig:simulation_half_wave_filter}c. To estimate the Rabi frequency and \ensuremath{T_\mathrm{1, ext}} just from the on-chip filter, we ground all the other co-calibrated ports except the qubit and the filter ports.

\begin{figure*}[ht]
\includegraphics[width=\linewidth]{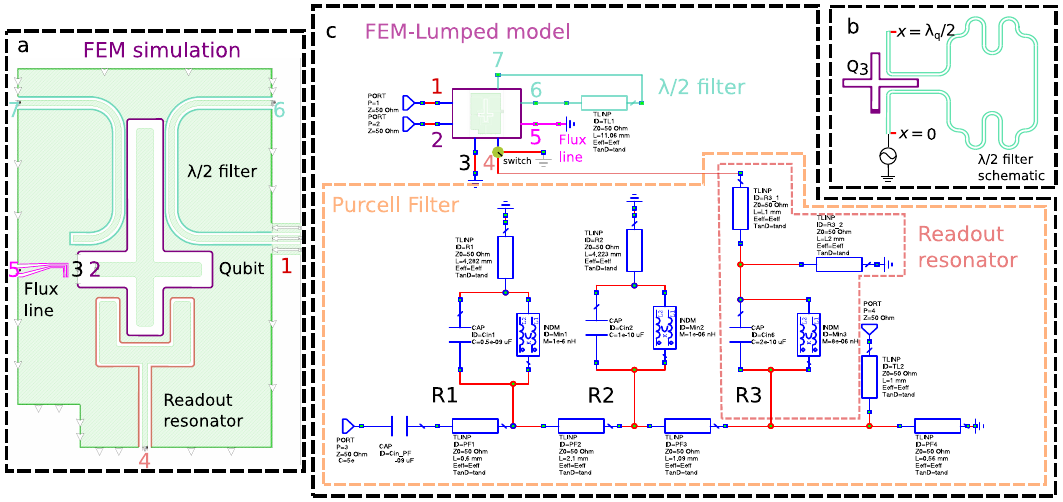}
\caption{\textbf{Full-circuit simulation scheme.} \textbf{a, b} (\textbf{a}) Simplified model of the \ensuremath{\lambda/2} filter of (\textbf{b}) is generated in Sonnet by removing the meanders and adding two co-calibrated ports 6 and 7 at both ends. Similarly, co-calibrated ports 2, 4, and 5 are added to the qubit island, readout resonator, and flux line, respectively in the Sonnet FEM simulation. Again, box-wall port 1 is used for the qubit drive through the \ensuremath{\lambda/2} filter. The FEM simulation for this geometry is carried out in the frequency range of 1--10~GHz, and the resulting 7-port S-parameter file is exported. \textbf{c} Full-circuit simulation scheme built in AWR utilizing the S-parameter file obtained from the FEM simulations, with the rest of the circuitry implemented with lumped elements. The meanders of the \ensuremath{\lambda/2} filter corresponding to a length of about 11~mm are added as a transmission line element between ports 6 and 7 of the S-parameter network. A Purcell filter together with three readout resonators R1, R2, and R3 is constructed with just the lumped components. The parameters of readout resonators and Purcell filters such as coupling capacitances, mutual inductances, and filter and resonator lengths are adjusted to match the designed values. Readout resonator R3 is coupled to the qubit Q3 at port 4 of the S-parameter network, while the flux line at port 5 is grounded. To estimate the Rabi frequency and the $T_\mathrm{1, ext}$ from just the \ensuremath{\lambda/2} filter, we also ground the resonator port, illustrated in (\textbf{a}) with a switch symbol. 
}
\label{fig:simulation_half_wave_filter}
\end{figure*}

To investigate the frequency splitting of the stopband observed in the characterization of the \ensuremath{\lambda/2} filter in the experiments, we simulate a lumped model of a \ensuremath{\lambda/2} filter that couples capacitively to qubit port at two separate locations, as depicted in Supplementary Fig.~\ref{fig:asymm_study}. We place a transmission line element between the coupling points, as shown in Supplementary Fig.~\ref{fig:asymm_study}a. We introduce a top and a bottom capacitor with capacitances \ensuremath{C^\mathrm{top}_\mathrm{d}} and \ensuremath{C^\mathrm{bot}_\mathrm{d}}, respectively, at the coupling points. By adding an asymmetry offset \ensuremath{\Delta C^\mathrm{top}_\mathrm{d}} to the top capacitor such that \ensuremath{C^\mathrm{top}_\mathrm{d} = C^\mathrm{bot}_\mathrm{d} - \Delta C^\mathrm{top}_\mathrm{d}}, we observe a frequency splitting of the stopband as a function of the offset relative to the bottom capacitor, as depicted in Supplementary Fig.~\ref{fig:asymm_study}b, c. The observed splitting of about 1~GHz in the experiment corresponds to a relative asymmetry offset of $4.9$\% as shown in Supplementary Fig.~\ref{fig:asymm_study}b, d. Furthermore, we conclude that the splitting of the stopband retains the simulated $T_\mathrm{1, ext}$ in the range of a few seconds; however, it does so with reduced bandwidth. This reduction in bandwidth occurs because the $\lambda/2$ filter no longer behaves as a high-order filter but as a first-order filter due to asymmetric coupling.

Additionally, we report the filter's sensitivity to the asymmetric offset of the top capacitor, which results in stopband splitting within the filter's bandwidth of approximately 500~MHz (estimated at $T_\mathrm{1, ext}$ of 1 ms for an ideal symmetric coupling case). Our simulations indicate that a deviation of \ensuremath{\Delta C^\mathrm{top}_\mathrm{d}/C^\mathrm{bot}_\mathrm{d}} within 1.2\% meets the filter sensitivity conditions.

\begin{figure*}[ht]
\includegraphics[width=\linewidth]{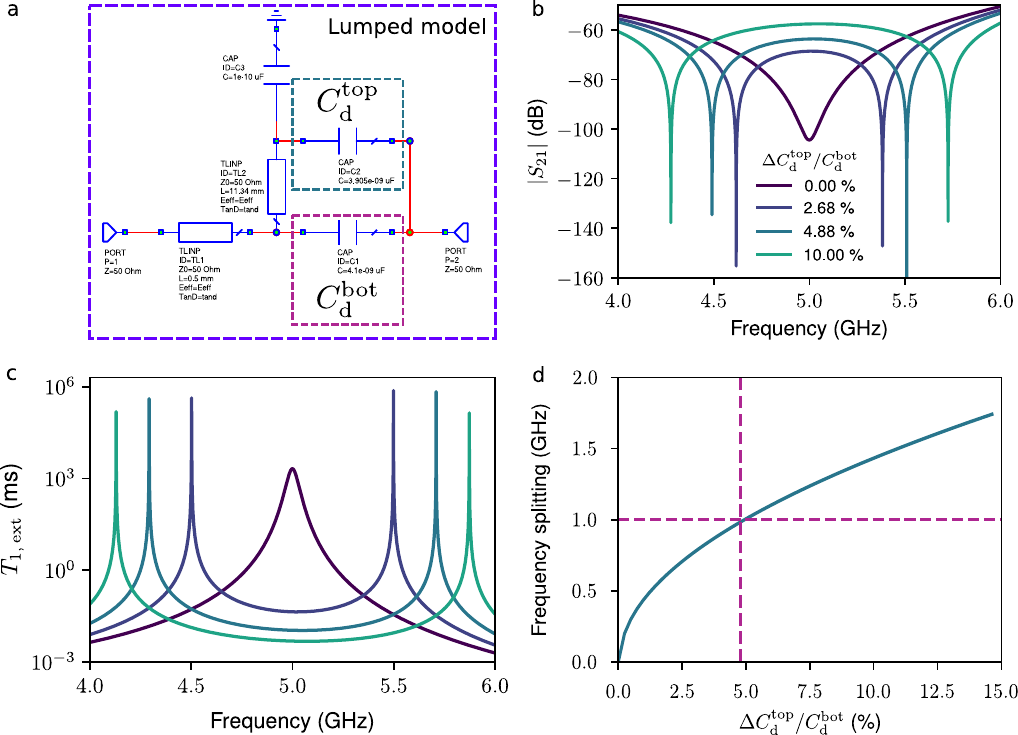}
\caption{\textbf{Stopband frequency splitting due to an asymmetric two-point coupling of the \ensuremath{\lambda/2} filter.} \textbf{a} Lumped model of the \ensuremath{\lambda/2} filter used for the simulation of asymmetric two-point coupling. An offset capacitance \ensuremath{\Delta C^\mathrm{top}_\mathrm{d}} is introduced to the top capacitor with capacitance \ensuremath{C^\mathrm{top}_\mathrm{d} = C^\mathrm{bot}_\mathrm{d} - \Delta C^\mathrm{top}_\mathrm{d}}. \textbf{b} Transmission amplitude from the qubit port to the drive line port as a function of frequency for the indicated asymmetry offset capacitances \ensuremath{\Delta C^\mathrm{top}_\mathrm{d}} relative to the bottom capacitor \ensuremath{C^\mathrm{bot}_\mathrm{d}}. \textbf{c} Simulated $T_\mathrm{1, ext}$ due to the external coupling as a function of frequency for similar asymmetric offset as in (\textbf{b}). \textbf{d} Frequency splitting of the stopband as a function of the  asymmetry offset obtained as the separation of the dips in the transmission amplitude in calculations identical to those exemplified in (\textbf{b}). For a relative asymmetry offset of approximately $4.9$\% (dashed purple vertical line) with respect to the bottom capacitance \ensuremath{C^\mathrm{bot}_\mathrm{d}}, we obtain the frequency splitting of roughly 1~GHz, equal to that observed in the experiment.
}
\label{fig:asymm_study}
\end{figure*}

\section*{Supplementary Note S2: Device fabrication}

A 200-nm-thick niobium film deposited on a six-inch high-resistivity silicon wafer is obtained from VTT Technical Research Centre of Finland. We spin-coat the wafer with a few-micron-thick optical resist AZ-5214E, and then dice the wafer into 33~mm~$\times$~33~mm squares using a Disco DAD3220 dicing saw. Following dicing, the dies are left overnight in acetone, then subjected to 10 minutes of sonication in acetone and a quick dip in isopropanol.

For patterning niobium capacitors, on-chip filters, resonators, and qubit islands, we utilize optical lithography with a Maskless Aligner MLA150 from Heidelberg. The process begins with HMDS priming of the die in the YES-3 prime oven. Next, we spin-coat the die with AZ-5214E resist using LabSpin6 at 4000 RPM for 40 seconds, achieving a resist thickness of roughly 1~\ensuremath{\mu\mathrm{m}}. This is immediately followed by a soft bake at 90 \ensuremath{^\circ\mathrm{C}} for 1.5 minutes.

A CAD file, featuring a 3-by-3 grid of 10~mm~$\times$~10~mm device geometries, is converted into a native job file for optical exposure of the die with MLA150. The mask aligner is set to operate with a 375-nm-wavelength laser diode source at a dose of 120~\ensuremath{\mathrm{mJ/cm^2}}. We then develop the resist in a 1:5 mixture of AZ-351B:DI~\ensuremath{\mathrm{H_2O}} for 50 seconds. Subsequently, the niobium layer is dry-etched using Plasmalab 80Plus from Oxford Instruments to reveal micron-size features. For reactive ion etching, we use SF6 (40~SCCM) and \ensuremath{\mathrm{O_2}} (20~SCCM) for six minutes. The die is left in acetone overnight for lift-off, followed by \ensuremath{\mathrm{O_2}} descum in the Plasmalab to remove any remaining resist residue.

For the junctions in the qubit, we fabricate Manhattan-style Josephson junctions in a SQUID form using electron beam lithography (EBL). We first spin-coat the die with MMA 8.5 EL13 polymer-based resist at 1550~RPM for a minute, followed by a post-bake at 160~\ensuremath{^\circ\mathrm{C}} for another minute. A layer of PMMA A4 resist is then added following similar steps. The resist is exposed to an electron beam using EBPG5200 from Raith, followed by development in a series of solvents: MIBK 1:3, methyl glycol 1:2, and isopropanol, each for 20 seconds. Aluminum is evaporated in a Plassys MEB700S2-III UHV at a rate of 0.2~nm/s to a thickness of 30~nm, then oxidized at a pressure of 3~mbar for seven minutes to create a tunnel barrier. This is followed by a second aluminum evaporation to a thickness of 40~nm. After overnight acetone soaking for lift-off, a second round of EBL and evaporation with similar parameters is carried out to establish contact between aluminum and niobium patterns. We dice a 33~mm~$\times$~33~mm die, yielding a total of nine devices from a single fabrication round. 

\section*{Supplementary Note S3: Experimental setup}

\begin{figure*}[ht]
\includegraphics[width=\linewidth]{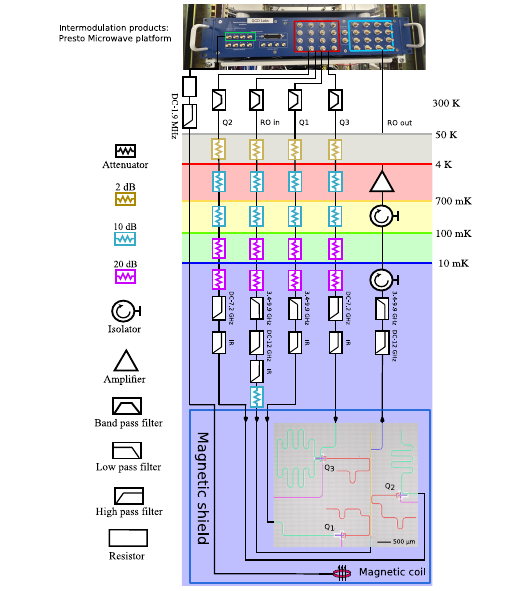}
\caption{\textbf{Schematic room temperature and cryogenic setup for the experiments.} The SMA-8 sample holder is not shown for clarity. The arrows heads on the lines indicate the main intended direction of propagation for the control and readout signals. The nominal temperatures of the different plates of the cryostat are shown on the right and the definitions of the circuit elements are shown on the left. 
}
\label{fig:exp_setup}
\end{figure*}

Following the fabrication of the device, we wire-bonded the device to a SMA-8 sample holder and placed the device inside a magnetic shield, which is installed at the base plate of a Bluefors XLD500 dilution refrigerator. Throughout the experiment, the refrigerator maintained a temperature range of 18--24~mK. An overview of the experimental setup is presented in Supplementary Fig.~\ref{fig:exp_setup}.  

We utilize three separate microwave lines to control the qubits Q1, Q2, and Q3 and a single line for readout (RO) input. Each control line has 62~dB and the RO input has 72~dB of attenuation just from the attenuators spread over multiple stages of the cryostat. For the readout output, we have employed two isolators and a high-electron-mobility transistor (HEMT) amplifier at the 4-K stage. In addition, low-pass and infrared (IR) filters are installed in all of the lines to block higher-frequency noise. Owing to the limited resource for direct-current (DC) lines, we use a magnetic coil that is installed on the sample holder to simultaneously bias all three qubits.

For the room temperature setup, we use the Presto microwave platform from Intermodulation Products for time-domain measurements and a vector network analyzer from Rohde \& Schwarz for continuous measurements, which we use in the initial characterizations of the Purcell filter and of the readout resonators. For the DC voltage source, we employ both Presto and the SIM900 from Stanford Research Systems, depending on the availability of SIM900 since the instrument generates stable DC voltage to bias the qubits.

\section*{Supplementary Note S4: Device characterization and measurement sequence}

\begin{figure*}[ht]
\includegraphics[width=\linewidth]{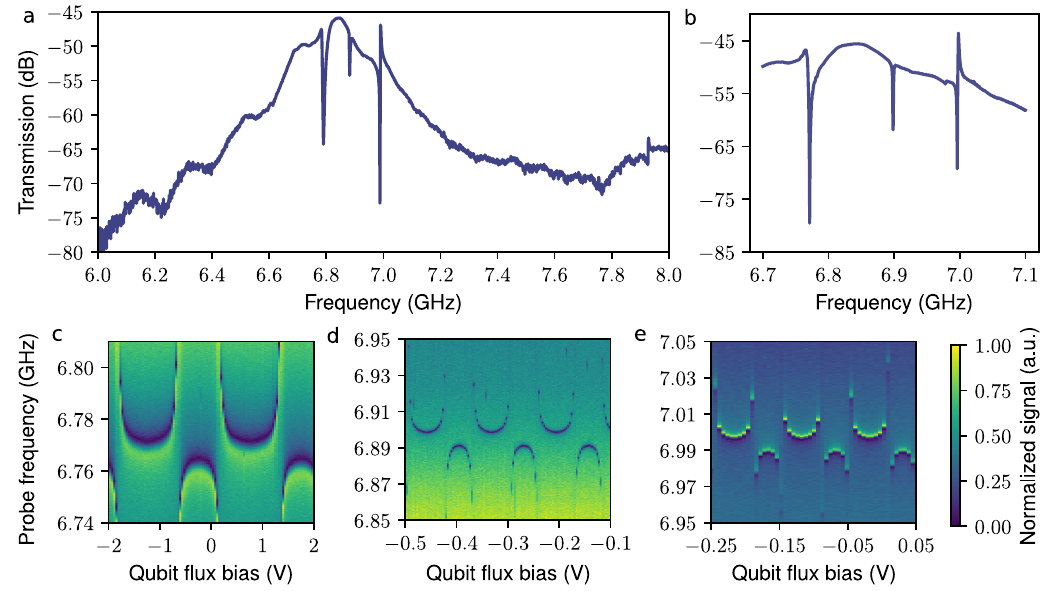}
\caption{\textbf{Characterization of Purcell filter and readout resonators using vector network analyzer.} \textbf{a,b} (\textbf{a}) Broadband transmission amplitude of a Purcell filter centered at 6.8~GHz and (\textbf{b}) a magnification on the sharp features corresponding to three readout resonators within the Purcell filter band as functions of probe frequency. \textbf{c,d,e}, Spectroscopy data similar to those in (\textbf{b}) of the readout resonators (\textbf{c}) Q1, (\textbf{d}) Q2, and (\textbf{e}) Q3 as functions of the flux bias voltage. 
}
\label{fig:Purcell_readout}
\end{figure*}

Immediately after the cool down, we implement continuous spectroscopic measurements to locate the readout resonator frequencies. In Supplementary Fig.~\ref{fig:Purcell_readout}a, b, we observe a broadband feature inherent to a Purcell filter centered at a frequency of 6.8~GHz. We find three sharp features corresponding to individual readout resonators within the Purcell filter band. In Supplementary Fig.~\ref{fig:Purcell_readout}c--e, we conduct resonator spectroscopy measurements for each qubit with varying flux bias voltage. The observed periodic pattern is typical of readout resonators with frequencies below the qubit frequency at the qubit sweet spot, leading to avoided crossings between the resonators and the qubits at a finite flux. 

In Supplementary Table~\ref{tab:device_char}, we summarize the parameters obtained from standard single-tone and two-tone time-domain measurements and from the initial characterization of our device. 

{\renewcommand{\arraystretch}{1.4}
\begin{table*}[ht]{\label{tab:device_char}}
\caption{Summary of measured device parameters. The asterisk (\textasteriskcentered{}) denotes a case where, owing to a very low \ensuremath{T_1} of qubit Q3 at the sweet spot, we estimate the dispersive shift from measured vacuum Rabi coupling strength \ensuremath{g}, detuning between the qubit and the readout resonator frequencies \ensuremath{\Delta = \omega_\mathrm{q} - \omega_\mathrm{r}}, and the anharmonicity of the qubit \ensuremath{\alpha} using \ensuremath{\chi = \alpha \frac{g^2}{\Delta (\Delta+\alpha)}}.}

\centering
\begin{tabular}{|p{3cm}|p{2cm}|p{3cm}|p{3cm}|p{3cm}|}
 \hline
 \textbf{Parameters } & \textbf{Symbols} & \textbf{Q1 (Standard)} & \textbf{Q2 ($\mathrm{\lambda/4}$ filter)} & \textbf{Q3 ($\mathrm{\lambda/2}$ filter)}\\
 \hline
Resonator frequency & \ensuremath{\mathrm{\omega}_\mathrm{r}/2\pi} & 6.8913 GHz & 6.9886 GHz & 6.7650 GHz\\
Resonator linewidth &  \ensuremath{\mathrm{\kappa}/2\pi} & 2.076 MHz & 2.805 MHz & 10.756 MHz\\
Qubit frequency & \ensuremath{\omega_\mathrm{q}/2\pi} & 7.773 GHz & 7.852 GHz & 7.640 GHz\\
Coupling strength &  \ensuremath{g/2\pi} & 72.455 MHz & 72.000 MHz & 68.750 MHz\\
Anharmonicity &  \ensuremath{\alpha/2\pi} & $-232.916$ MHz & $-229.847$ MHz & $-234.520$ MHz\\
Dispersive shift & \ensuremath{\chi/2\pi} & $-2.300$ MHz & $-2.025$ MHz & $-1.978$ MHz (\textasteriskcentered{})\\
Relaxation time & \ensuremath{T_1} & 7.446 \ensuremath{\mu}s & 0.481 \ensuremath{\mu}s & 0.06 \ensuremath{\mu}s \\
Dephasing time & \ensuremath{T^\mathrm{*}_\mathrm{2}} & 8.289 \ensuremath{\mu}s & 0.929 \ensuremath{\mu}s & --- \\
Hahn echo time & \ensuremath{T^\mathrm{E}_\mathrm{2}} & 8.475 \ensuremath{\mu}s & 0.934 \ensuremath{\mu}s & --- \\
 \hline
\end{tabular}
\end{table*}
}

\begin{figure*}[ht]
\includegraphics[width=\linewidth]{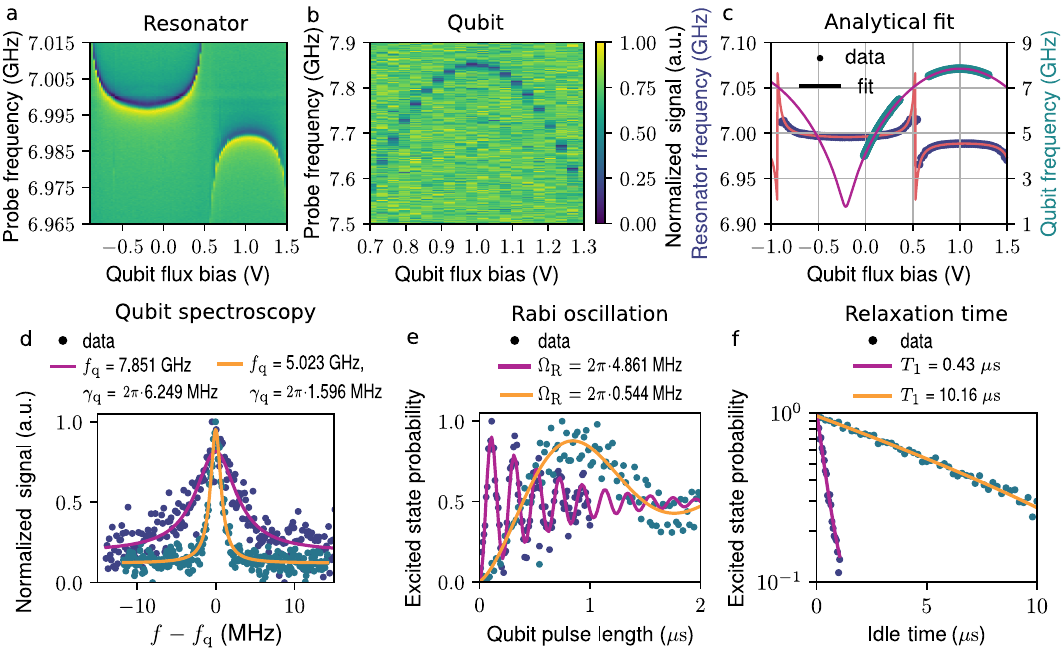}
\caption{\textbf{Measurement sequence for on-chip-filter characterization for qubit Q2.} \textbf{a,b} Normalized readout signal for the spectroscopy measurements for (\textbf{a}) the readout resonator and (\textbf{b}) the qubit Q2 as a function of the flux bias voltage and the probe frequency. \textbf{c} Measured resonator frequency (blue marker) and qubit frequency (turquoise marker) as a function of the bias voltage, obtained from experiments similar to those of (\textbf{a}) and (\textbf{b}), respectively. Analytical fit of the resonator frequency (red solid) and qubit frequency (purple solid) to the measured data. \textbf{d,e,f}~For a fixed pulse amplitude, we implement (\textbf{d}) qubit spectroscopy by sweeping the probe frequency $f$ around qubit frequency $f_\mathrm{q}$, (\textbf{e}) Rabi oscillation by varying the qubit pulse length, and (\textbf{f}) \ensuremath{T_1} measurement by changing an idle delay between the \ensuremath{\pi} pulse of the qubit and the readout pulse. The measured (markers) data and corresponding fit (solid line) at two different flux bias voltages in (\textbf{d})--(\textbf{f}) correspond to the sweet spot of the qubit at around 7.85~GHz (blue markers, purple line) and to the filter frequency around 5~GHz (turquoise markers, orange line). The fitting parameters are shown on top of the corresponding figures.
}
\label{fig:meas_sequence}
\end{figure*}

We show a typical measurement sequence for the characterizion of the on-chip filters in Supplementary Fig.~\ref{fig:meas_sequence} for qubit Q2 coupled to the \ensuremath{\lambda/4} filter. We begin by determining the flux-bias-voltage-dependent frequencies of the resonator and of the qubit through the single-tone and two-tone spectroscopy measurements, respectively, as depicted in Supplementary Fig.~\ref{fig:meas_sequence}a,~b. We fit the resonator and the qubit spectroscopic data to the analytical model of Ref.~\cite{Fedorov2019}, as shown in Supplementary Fig.~\ref{fig:meas_sequence}c. With this fit, we can automate the search of the resonator and the qubit frequencies in large frequency ranges for a given bias voltage.

In Supplementary Fig.~\ref{fig:meas_sequence}d-e, we show a measurement sequence for two different qubit flux bias voltages, corresponding to the qubit frequencies of 7.851~GHz at the sweet-spot and 5.023~GHz at the $\lambda/4$ filter stop band. To measure the Rabi frequency and the qubit \ensuremath{T_1}, we first determine the DC voltage for the desired qubit flux bias according to Supplementary Fig.~\ref{fig:meas_sequence}c. Subsequently, we set the correct readout frequency according to the fit in Supplementary Fig.~\ref{fig:meas_sequence}c. Next, we again carry out qubit spectroscopy at very low power, fitting the data to a Lorentzian curve to accurately determine the qubit frequency, see Supplementary Fig.~\ref{fig:meas_sequence}d. 

Using the carefully extracted qubit frequency, we implemented Rabi oscillations and fit the data to a decaying cosine function to determine the \ensuremath{\pi} pulse length. The Rabi oscillation measurements are conducted at fixed pulse voltage of 0.05~V, as shown in Supplementary Fig.~\ref{fig:meas_sequence}e. We then measure the \ensuremath{T_1} time by applying a \ensuremath{\pi} pulse, as obtained from the Rabi oscillation fit, and by varying the delay before the final readout pulse. We fit the measured data to a decaying exponential function to obtain the \ensuremath{T_1} time and its uncertainty, see Supplementary Fig.~\ref{fig:meas_sequence}f. The entire measurement sequence is automated, requiring only the bias voltage as an input parameter.

\section*{Supplementary Note S5: Observations and discussion on \ensuremath{T_1}}

\begin{figure*}[ht]
\includegraphics[width=\linewidth]{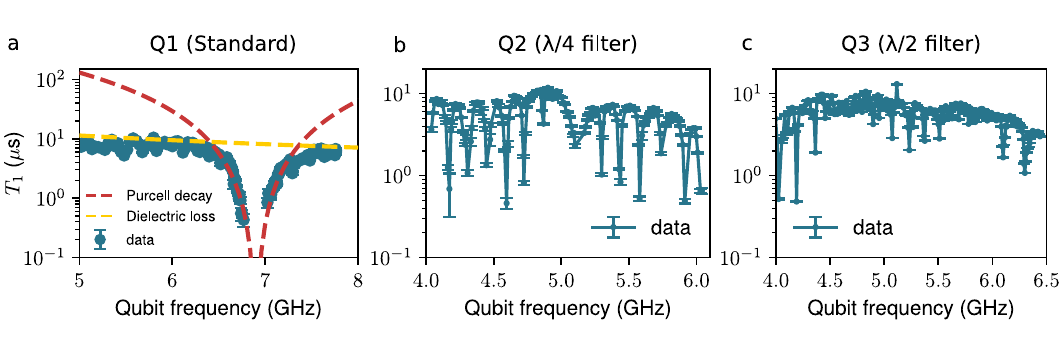}
\caption{\textbf{Energy decay time $T_1$ of each qubit at different frequencies.} \textbf{a} Measured $T_1$ of qubit Q1 (markers) and corresponding results of fitted models (dashed lines), Purcell decay (red) and dielectric-loss (yellow), as functions of qubit frequency. From the fit to the dielectric-loss model, we obtain a loss tangent of $3 \times 10^{-6}$, manifesting as the dominant loss mechanism in our devices far from the resonance of the qubit and its readout resonator. \textbf{b,c} Measured \ensuremath{T_1} of qubit (\textbf{b}) Q2 and (\textbf{c}) Q3 with an additional off-chip low-pass filter with a cut-off frequency at around 3~GHz installed in qubit control line. The error bars represent \ensuremath{1\sigma} fitting uncertainty.
}
\label{fig:Dielectric_LPF_study}
\end{figure*}

We observe a maximum measured \ensuremath{T_1} of around 10~$\mu\mathrm{s}$ across all qubits. To investigate the reason behind this bound, we measure the \ensuremath{T_1} of qubit Q1 in the vicinity of its resonator frequency. We fit the measured data to the Purcell decay and the dielectric-loss models, as shown in Supplementary Fig.~\ref{fig:Dielectric_LPF_study}a. For the dielectric-loss model, we follow a similar approach used for the Fluxonium and the Unimon qubit~\cite{Zhang2021, Hyyppa2022}. However for the transmon qubit, we assume the capacitive loss to be dominant and utilize the relation
\begin{align}
    1/T_{1, \mathrm{diel}} &= \frac{16E_\mathrm{C}}{\hbar Q_\mathrm{cap}} \coth\bigg({\frac{\hbar\omega_\mathrm{q}}{2k_\mathrm{B}T}}\bigg)\big|\bra{0}\hat{n}\ket{1}\big|^2\\
     &= \frac{\sqrt{8E_\mathrm{J}E_\mathrm{C}}}{\hbar Q_\mathrm{cap}}\coth\bigg({\frac{\hbar\omega_\mathrm{q}}{2k_\mathrm{B}T}}\bigg)\label{eq:cap_dl},
\end{align}
where \ensuremath{\hbar} is the reduced Planck constant, \ensuremath{k_\mathrm{B}} is the Boltzmann constant, \ensuremath{\omega_\mathrm{q}} is the angular frequency of the qubit, \ensuremath{E_\mathrm{J}} is the Josephson energy, \ensuremath{E_\mathrm{C}} is the charging energy, $T$ is the temperature of the dissipative environment, $\hat{n} = \textrm{i}\big(\frac{E_\mathrm{J}}{8E_\mathrm{C}}\big)^{\frac{1}{4}}(\hat{b}^\dagger - \hat{b})$ is the number operator of the qubit with $\hat{b}$ and $\hat{b}^\dagger$ as the annihilation and creation operators. The ket vectors $\ket{0}$ and $\ket{1}$ represents the ground and the excited state of the qubit, respectively. 

We fit the equation \eqref{eq:cap_dl} to the measured data to obtain a realistic estimate for the loss tangent $\tan\delta = 1/Q_\mathrm{cap}$. From the fit, we obtain a value of $\tan\delta = 3 \times 10^{-6}$, suggesting that dielectric loss is the primary contributor to the observed \ensuremath{T_1} bound. For the Purcell decay model, we employ the relation \ensuremath{1/T_\mathrm{1, Purcell} = \kappa g^2/|\omega_\mathrm{q} - \omega_\mathrm{r}|^2}, where $g$ is the coupling strength between the qubit and the resonator, $\kappa$ is the linewidth of the resonator, and $\omega_\mathrm{r}$ is the angular frequency of the resonator. 

Furthermore, following a similar approach to the one used in Ref.~\cite{Xia2023}, we show that the use of an off-chip low-pass filter in the qubit control line, facilitating a passband at the subharmonic frequency while introducing a stopband at the qubit frequency, creates unwanted resonances in the qubit drive line that modulate the measured \ensuremath{T_1} in extreme ranges, as depicted in Supplementary Fig.~\ref{fig:Dielectric_LPF_study}b, c. By employing the integrated on-chip filters, we are not faced with such an issue and the resulting engineering and scaling challenge.


\end{document}